\documentclass[review]{elsarticle}

\usepackage{lineno,hyperref}
\modulolinenumbers[5]

\usepackage{url}  
\usepackage{graphicx} 
\usepackage{revsymb}
\usepackage{bm}
\usepackage{textcomp}
\usepackage{revsymb}
\usepackage{amssymb}
\usepackage{ifsym}
\usepackage{wasysym}
\usepackage{txfonts}
\usepackage{hyperref}
\usepackage{tipa}

\journal{arXiv}

\begin{document}

\begin{frontmatter}









\title{High-temperature instability of artificial cuprorivaite: \\ a study using thermal analysis, X-ray powder diffractometry and polarized light microscopy}




\author[GPI]{Vladimir~A.~Yuryev\corref{Yuryev}}
\cortext[Yuryev]{Corresponding author}
\ead{vyuryev@kapella.gpi.ru}

\author[GPI]{Sergey~V.~Kuznetsov}
\ead{kouznetzovsv@gmail.com}

\author[GPI]{Alexander~A.~Alexandrov}
\ead{alexandrov1996@yandex.ru}

\author[GOSNIIR]{Tatyana~V.~Yuryeva} 
\ead{yuryevatv@gosniir.ru}


\address[GPI]{A.\,M.\,Prokhorov General Physics Institute of the Russian Academy of Sciences,\\38 Vavilov Street, 119991 Moscow, Russia}

\address[GOSNIIR]{The State Research Institute for Restoration, Building 1, 44 Gastello Street, Moscow 107114, Russia}






\begin{abstract}
	CaCuSi$_4$O$_{10}$ powder was studied by differential scanning calorimetry and thermogravi\-metry methods in the  range from room temperature to 1450\,{\textcelsius} at heating and cooling rates of 20\,{\textcelsius}/min.
	The process of decomposition of cuprorivaite, the composition and  transformations of its decomposition products during successive heat treatments were also studied by powder X-ray diffraction and polarization optical microscopy techniques.
	It was found that CaCuSi$_4$O$_{10}$ starts to decompose by incongruent melting at a temperature of about 1020\,{\textcelsius}, with the minimum of the endothermic DSC peak associated with this process being at 1064.4\,{\textcelsius};
	CaCuSi$_4$O$_{10}$ decomposes irreversibly and subsequent cyclic annealings up to a temperature of 1450\,{\textcelsius} at heating and cooling rates of 20\,{\textcelsius}/min do not cause its re-synthesis;
	CaCuSi$_4$O$_{10}$ transforms into a two-phase system consisting of acicular crystals of monoclinic tridymite fused with green glass with the composition CuO\,--\,Cu$_2$O\,--\,CaO\,--\,SiO$_2$, with the weight ratio of tridymite to glass being about $12:13$, as a result of two successive annealings up to the temperature of 1450\,{\textcelsius}.
\end{abstract}

\begin{graphicalabstract}
	\includegraphics[width=\textwidth]{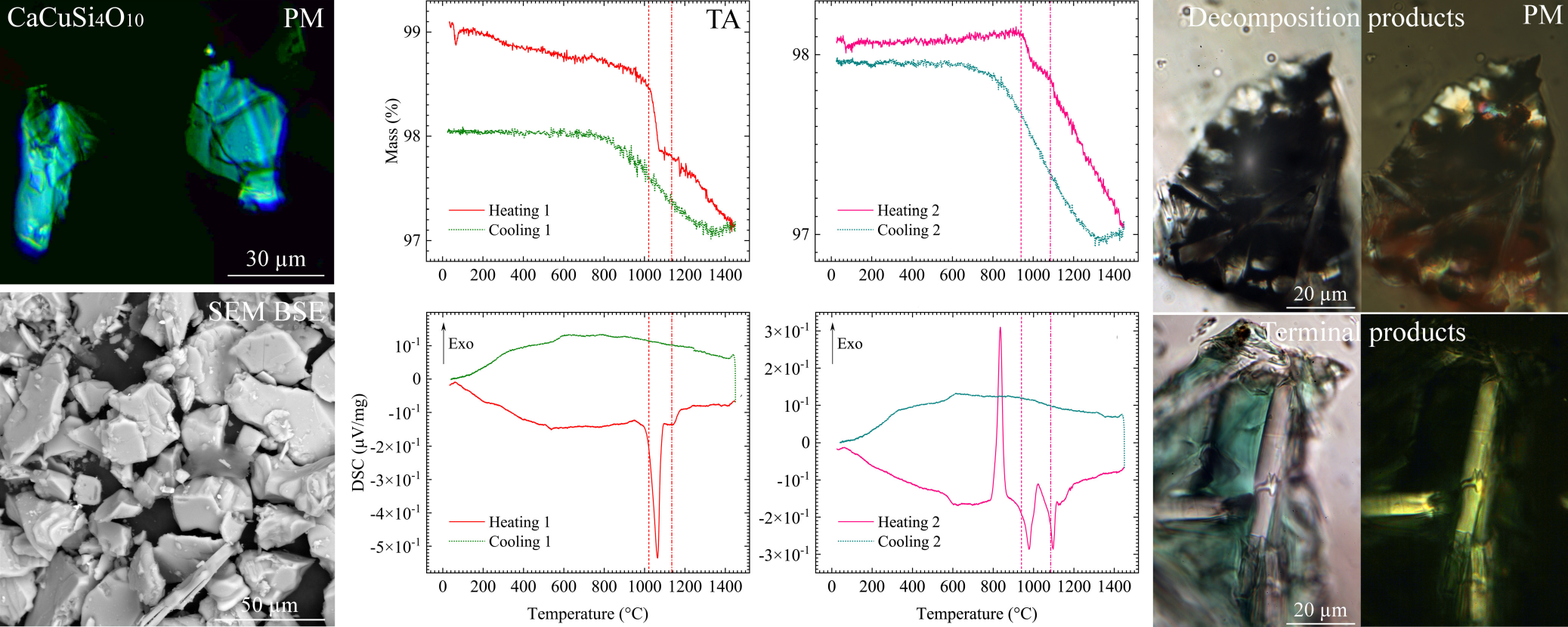}
\end{graphicalabstract}

\begin{highlights}
	\item 
	CaCuSi$_4$O$_{10}$ is studied by thermal analysis, X-ray diffraction and polarizing microscopy
	\item 
	CaCuSi$_4$O$_{10}$ starts decomposing by incongruent melting at a temperature of about 1020\,{\textcelsius}
	\item 
	CaCuSi$_4$O$_{10}$ decomposes irreversibly, heating to 1450\,{\textcelsius} does not lead to its re-synthesis
	\item 
	Terminal products of CaCuSi$_4$O$_{10}$ cyclic annealings are tridymite and CuO\,--\,Cu$_2$O\,--\,CaO\,--\,SiO$_2$ glass 
	\item 
	Production process temperature of structures based on CaCuSi$_4$O$_{10}$ must not exceed 1020\,{\textcelsius}
\end{highlights}

\begin{keyword}
\texttt{calcium copper phyllo-tetrasilicate 
	\sep Egyptian blue
	\sep thermal decomposition
	\sep differential scanning calorimetry
	\sep thermogravimetry
	\sep X-ray diffractometry
	\sep polarizing microscopy
	}
\end{keyword}

\end{frontmatter}


\newpage

\section{Introduction}\label{sec:intro}

Artificial cuprorivaite (calcium copper phyllo-tetrasilicate CaCuSi$_4$O$_{10}$) \cite{Cuprorivaite, Cuprorivaite_mindata} is commonly believed to be the earliest synthetic pigment invented by mankind  \cite{EB_&_UB, Egyptian_Blue_book,EB_Europe_Egypt-1910, EB_technology, EB_production_technology_Mesopotamia, Egyptian_Blue_Persia}.
Due to its provenance in Ancient Egypt during the 4th Dynasty of the pharaohs (cca.~2613--2494 BC), this substance is usually referred to as Egyptian blue or, somewhat less commonly, Alexandria frit \cite{Egyptian_Blue_book,EB_Europe_Egypt-1910, Pigment_Compendium};
previously, this pigment was sometimes called Pompeian blue \cite{Lefranc-1930, CaCuSi4O10_Falk}. 
It was widespread in most countries throughout Western Asia and the Mediterranean Region, such as Mesopotamia, Persia, Assyria, Urartu, Parthia and Greece, in the antiquity 
\cite{Egyptian_Blue_book, EB_Europe_Egypt-1910, EB_technology, EB_production_technology_Mesopotamia, Egyptian_Blue_Persia, Pigment_Compendium, Ancient_Blue_Purple_Pigments, EB_Egypt_Aegean+Near_East, EB@Kos_article, Grenberg_&_Pisareva-Erebuni_1982, Pisareva-Erebuni_1987, EB_Urartu, EB_Turkey_Ayanis, EB_Turkey_Lake_Van, Egyptian_Blue_Visible-Induced, Egyptian_Blue_Fayum_Portraits-2015, Egyptian_Blue_Fayum_Portraits-2018, Old_Nisa_Veresotskaya}.
Later, Egyptian blue became known throughout Roman Europe and even beyond, and was used until the 3rd century AD 
\cite{EB_Norway, Micro-EB_Arch_England}.

Despite the widespread belief that Egyptian blue was forgotten in the Middle Ages \cite{Egyptian_Blue_book},
it was unexpectedly discovered in early medieval frescoes \cite{EB_Zn-rich_Nicola2019,EB_Zn-rich_Castelseprio_NICOLA2018465,EB_Spain}
as well as
some works of art from the Renaissance, such as the painting ``Saint Margaret'' by Ortolano Ferrarese \cite{Egyptian_Blue_Benvenuto-1524} or the fresco ``The Triumph of Galatea'' by Raphael \cite{EB_Raphael}.

Surprisingly, Egyptian blue has lately been detected in the  painting ``Birch.  Spring'' made by Robert Falk, one of the prominent artists of Russian avant-garde, in 1907  
\cite{CaCuSi4O10_Falk}.

The above brief excursion into the history of Egyptian blue presents this compound solely as a historical art pigment.
However, at present, artificial cuprorivaite is gradually attracting increasing attention from researchers as a promising material for various branches of photonics and electronics \cite{Egyptian_Blue_New_Life,EB_Fluorescent_Nanosheets,EB_solar_cells,EV_battery_electrodes,EB_nanosheets_biophotonics,EB_bioimaging,EB_Bon_Regeneration,EB_fingerprint_powder,EB_methanol_fingerprints,EB_latent_fingermarks}, 
due to its layered structure, which enables obtaining thin, down to monolayer thick, nanosheets 
\cite{EB_ML_structure&properties,Egyptian_Blue_Nano}, 
as well as owing to highly efficient photoluminescence in the near IR range with a peak wavelength of around 900 nm
\cite{PL_EB_HB_HP,Egyptian_Blue_PL,MCuSi4O10_IR_phosphors,T-dependent_NIR_emission,EB_Magnetic_Optical, EB_Photo@Cathodo}. 
In particular, the literature discusses possible applications of this substance in such diverse fields as energy harvesting and storage, medical and biological photonics, and forensics \cite{Egyptian_Blue_New_Life,EB_Fluorescent_Nanosheets,EB_solar_cells,EV_battery_electrodes,EB_nanosheets_biophotonics,EB_bioimaging,EB_Bon_Regeneration,EB_fingerprint_powder,EB_methanol_fingerprints,EB_latent_fingermarks}.
Currently, most studies investigate the properties and potential applications of powders consisting of cuprorivaite scales with transverse dimensions of units or tens of micrometers.
We believe, however, that single-crystalline CaCuSi$_4$O$_{10}$ \cite{EB_single-crystal} is of no less interest for future applications, e.g., for laser technology, with thin films being the most promising.
The synthesis of bulk single crystals and thin films of cuprorivaite, as well as the formation of structures based on them, requires knowledge of the upper temperature limit of stability of this compound.

This article is devoted to the study of this issue using the thermal analysis technique.
In addition, X-ray powder diffraction and polarization optical microscopy methods were used in the work to study the process of decomposition of cuprorivaite, the composition as well as chemical and phase transformations of its decomposition products during successive heat treatments.

\section{Sample preparation, experimental methods and equipment} \label{sec:experiment}

\subsection{Samples } \label{subsec:samples}

\begin{figure}[htb] 	
			\includegraphics[width=0.7\textwidth]{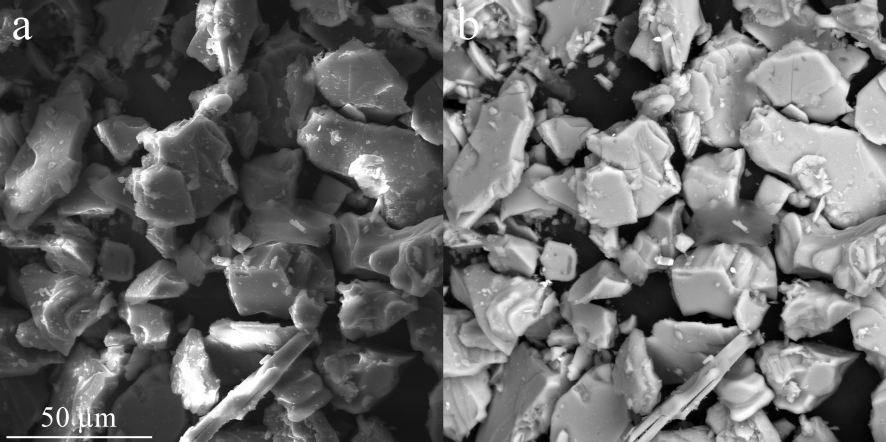}\\   
			\includegraphics[width=0.7\textwidth]{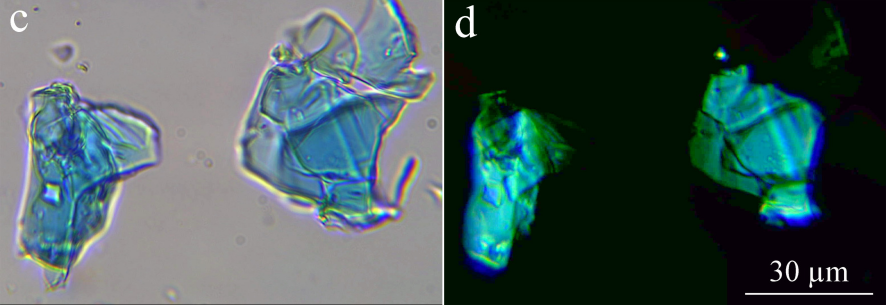}     
		\caption{\label{fig:SEM+PM} 
			Complementary pairs of SEM SE (a) and BSE (b) images of a sample of Egyptian blue (Kremer \#\,10060), and of polarized-light micrographs of two flakes of the CaCuSi$_4$O$_{10}$ powder obtained using  parallel~(c) and crossed~(d) polars.
		}
	\end{figure}

Powder of commercial Egyptian blue pigment (Kremer \#\,10060) was utilized to prepare samples of artificial cuprorivaite (calcium copper phyllo-tetrasilicate CaCuSi$_4$O$_{10}$) \cite{Cuprorivaite, Egyptian_Blue_book} for the investigations.
Since we had thoroughly analyzed this substance previously and verified its high purity 
(no foreign impurities as well as intentional admixtures had been detected)
\cite{CaCuSi4O10_Falk, EB_Photo@Cathodo}, 
no additional treatment was done before the experiments.
Scanning electron microscopy and polarization microscopy confirmed the single-phase composition of the samples (Fig.~\ref{fig:SEM+PM}).


\subsection{Methods and instruments} \label{subsec:Method}


Thermal analysis was carried out using an STA\,449\,F3 Jupiter (NETZSCH) simultaneous thermal analyzer; 
differential scanning calorimetry (DSC) and thermogravimetry (TG) analyzes were performed simultaneously.
DSC--TG curves were recorded in a flow of a mixture of synthetic air (N$_2\,+$\,O$_2$, $80/20$~vol.\%) at a rate of 50~ml$/$min and argon as protective gas in heater chamber at a rate of 50~ml$/$min.
Samples were heated up to the temperature of 1450\,{\textcelsius} and then cooled down to room temperature at a heating and cooling rate of 20\,{\textcelsius}$/$min in a Pt furnace
PtRh20 crucibles with a volume of 85~{\textmu}l and a lid were used.

\begin{figure}[htb]
	\begin{minipage}[l]{1\textwidth}{\vspace{-0cm}}{\hspace{-2.5cm}}
					\begin{tabular}{ll}
	\includegraphics[height=5cm]{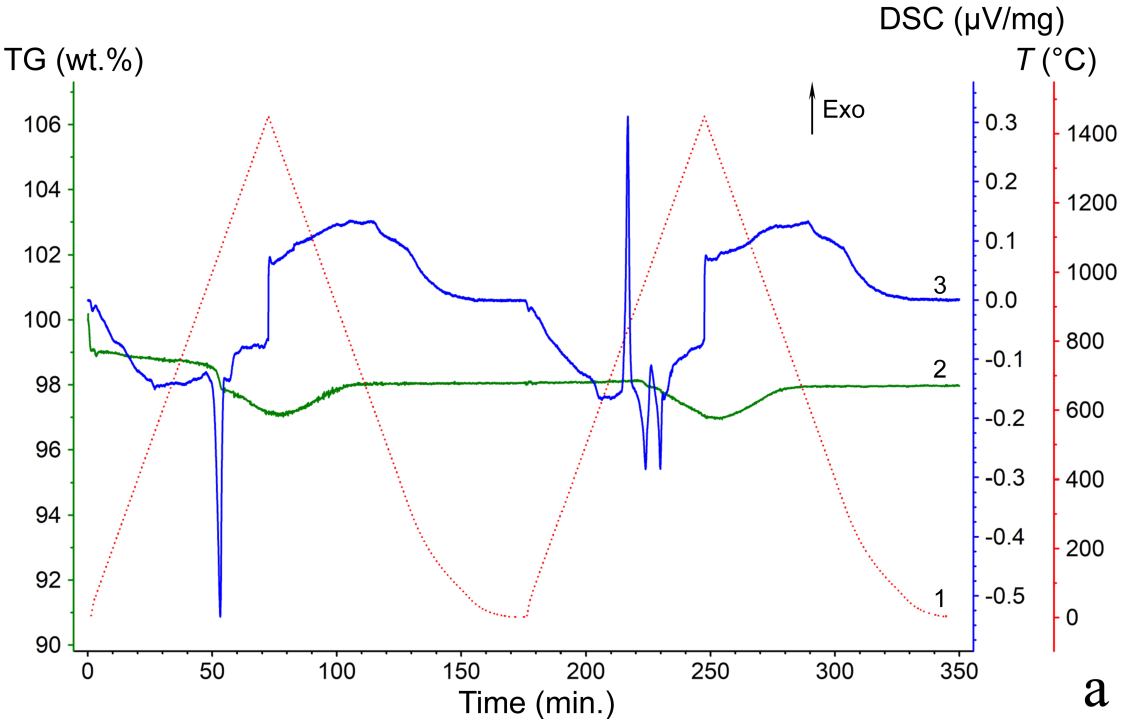}  &
	\includegraphics[height=5cm]{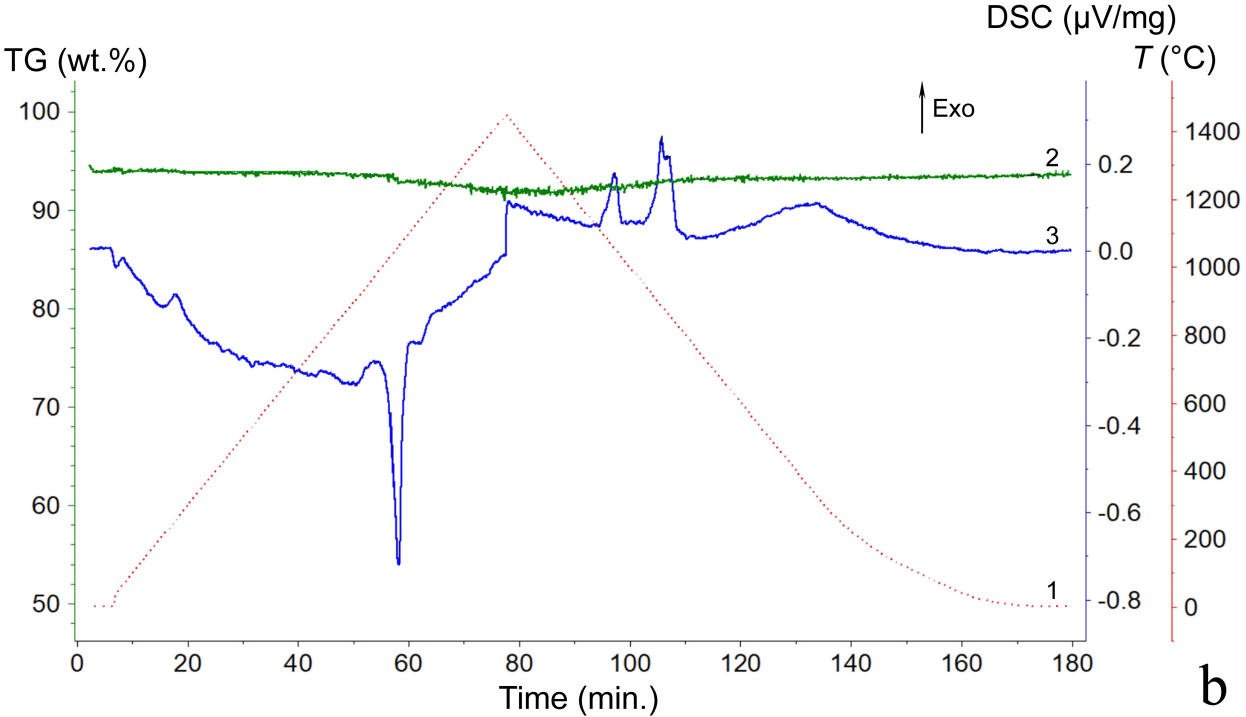} \\
					\end{tabular} 
	\end{minipage}
	\caption{\label{fig:DSC+TGA} 
		Thermal analysis curves: 
		(a) a double-cycle experiment;
		(b) a single-cycle experiment;
		(1) temperature cyclograms, red dotted lines and $y$-axes; 
		(2) TG curves, green lines and $y$-axes; 
		(3) DSC curves, blue lines and $y$-axes.
	}
\end{figure}

There were two types of experiments:
one type consisted of 
two successive similar cycles of heating from room temperature to 1450~{\textcelsius} and cooling to room temperature (Fig. ~\ref{fig:DSC+TGA}\,a, curve~1), followed by X-ray phase analysis;
the other included 
a single cycle of heating and cooling in the same conditions as the first one (Fig.~\ref{fig:DSC+TGA}\,b, curve~1), also followed by X-ray phase analysis.
In the first type of experiment, 
the sample mass ($m_{\rm s}$) was 13.0~mg, the mass of the crucible with the sample and the mass of the reference one were 241.4 and 243.3~mg, respectively,
whilst in the second type of experiment, 
the sample mass was 4.0~mg, the mass of the crucible with the sample and the mass of the reference one were 237.5 and 239.3~mg, respectively.


Phase analysis of the samples was carried out after each experiment by means of X-ray powder diffraction (XRD) \cite{Powder_XRD}.
Powder patterns were acquired using a D8 Advance diffractometer (Bruker); 
an X-ray source operating at Cu~K$_{\alpha}$ band was used. 
XRD patterns were recorded in air over the $2\theta$ range from 7 to 80$^{\circ}$ with a step of 0.02$^{\circ}$; 
the signal collection time was 0.5~s$/$step. 
The samples were rotated at a frequency of 20 rpm.
Patterns were indexed using the PDF-2 and COD \cite{COD_presentation-2} databases.


A Mira 3~XMU (Tescan Orsay Holding) scanning electron microscope (SEM) was employed for sample imaging.
Complementary pairs of SEM images were recorded simultaneously by registering secondary electrons (SE) and backscattered electrons (BSE).


Light micrographs were obtained using a POLAM L213-M (LOMO) polarizing microscope.

Specimens for polarization light microscopy (PM) were prepared as follows.
Small pieces were broken off from the samples obtained after thermal analysis and thoroughly ground.
Then the technique of making permanent slides, widely used in petrography and adopted for pigment analysis by specialists studying cultural heritage sites, was applied.
This involved placing particles of a sample on a microscope slide, adding a drop of Canada balsam in orthoxylene, and securing the prepared slide with a coverslip.
The permanent preparations were examined using the microscope with parallel and crossed polars \cite{Artists_Pigments_book_PM, Polarization_Microscopy, Optical_PM-Microscopy, Optical_Microscopy-PM}.

\section{Results and their interpretation } \label{sec:results}

\subsection{Thermal analysis} \label{subsec:thermal}

\subsubsection{Double-cycle experiment} \label{subsubsec:double}

\begin{figure}[t]
	\begin{minipage}[l]{1\textwidth}{\vspace{-0.0cm}}{\hspace{-0cm}}
				\begin{tabular}{rr}
	\includegraphics[height=4cm]{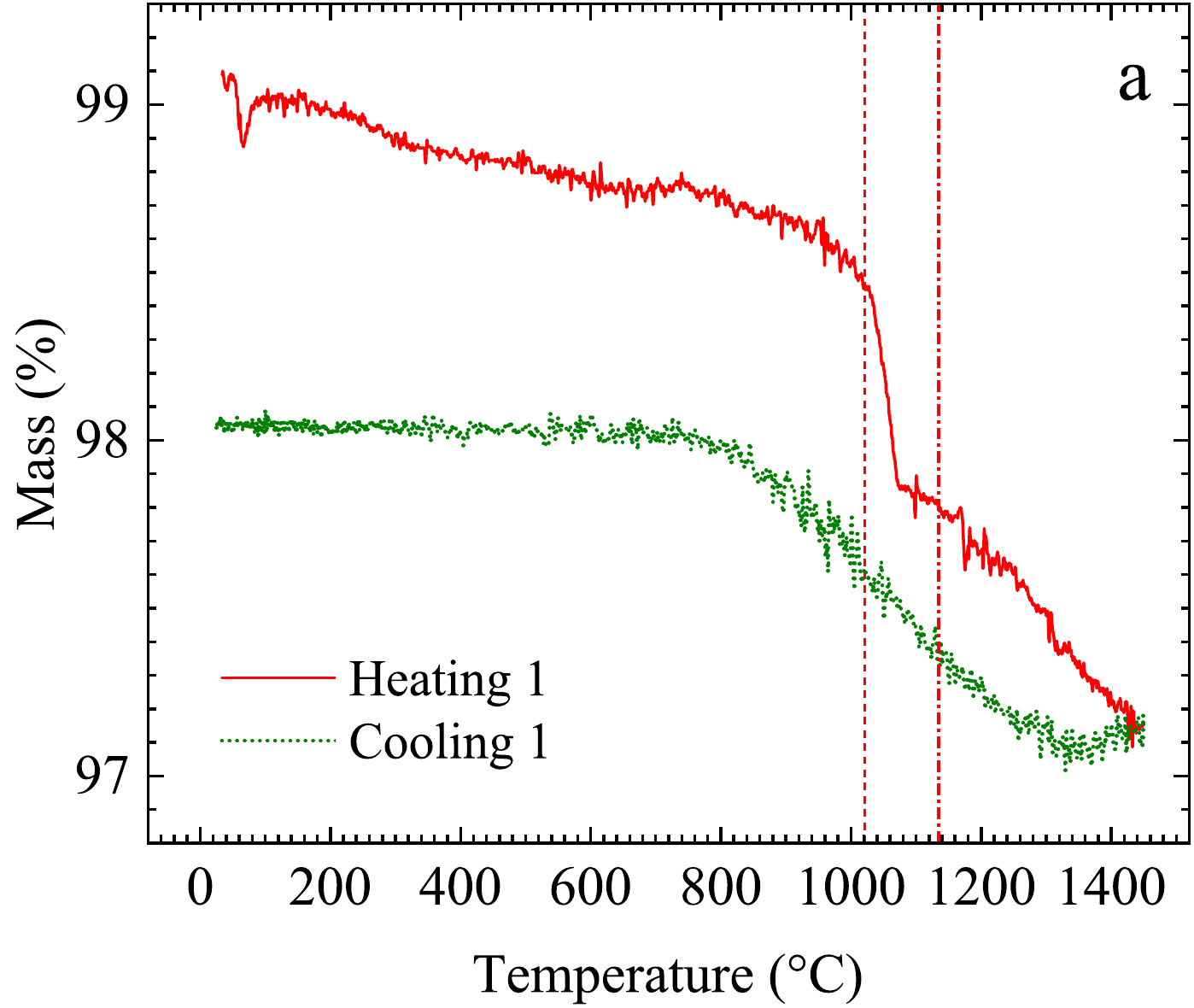} &
	\includegraphics[height=4cm]{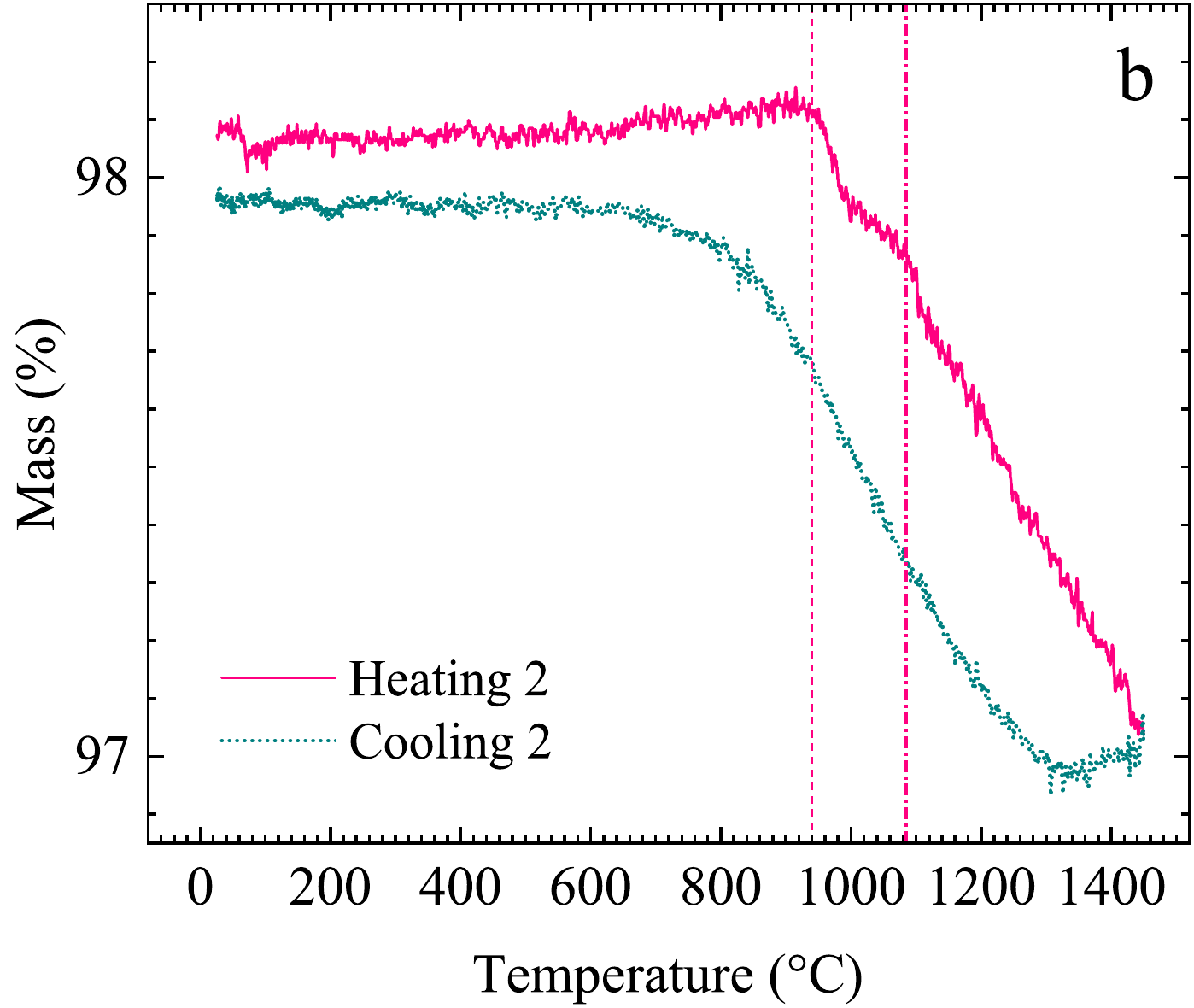}\\
	\includegraphics[height=4cm]{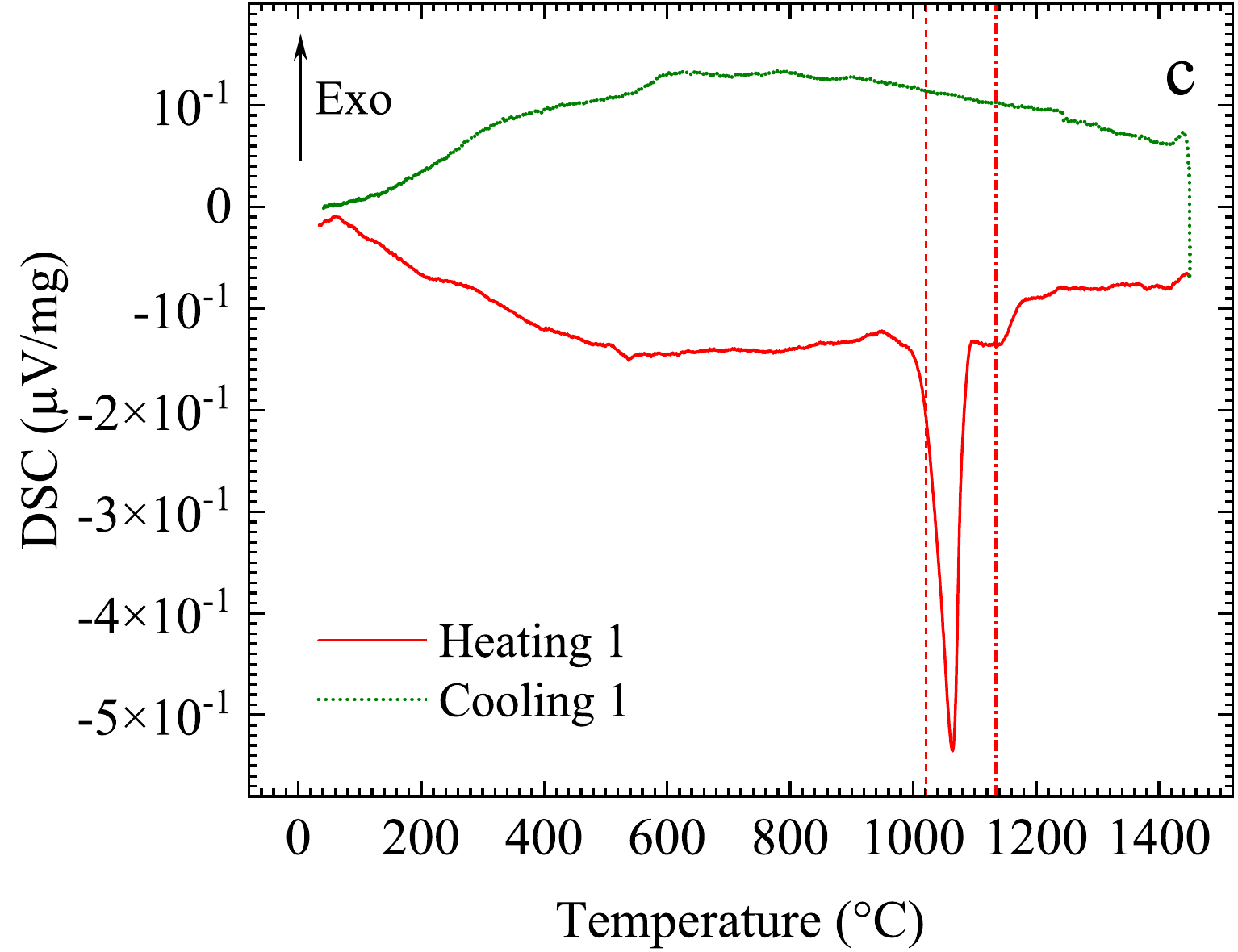}&
	\includegraphics[height=4cm]{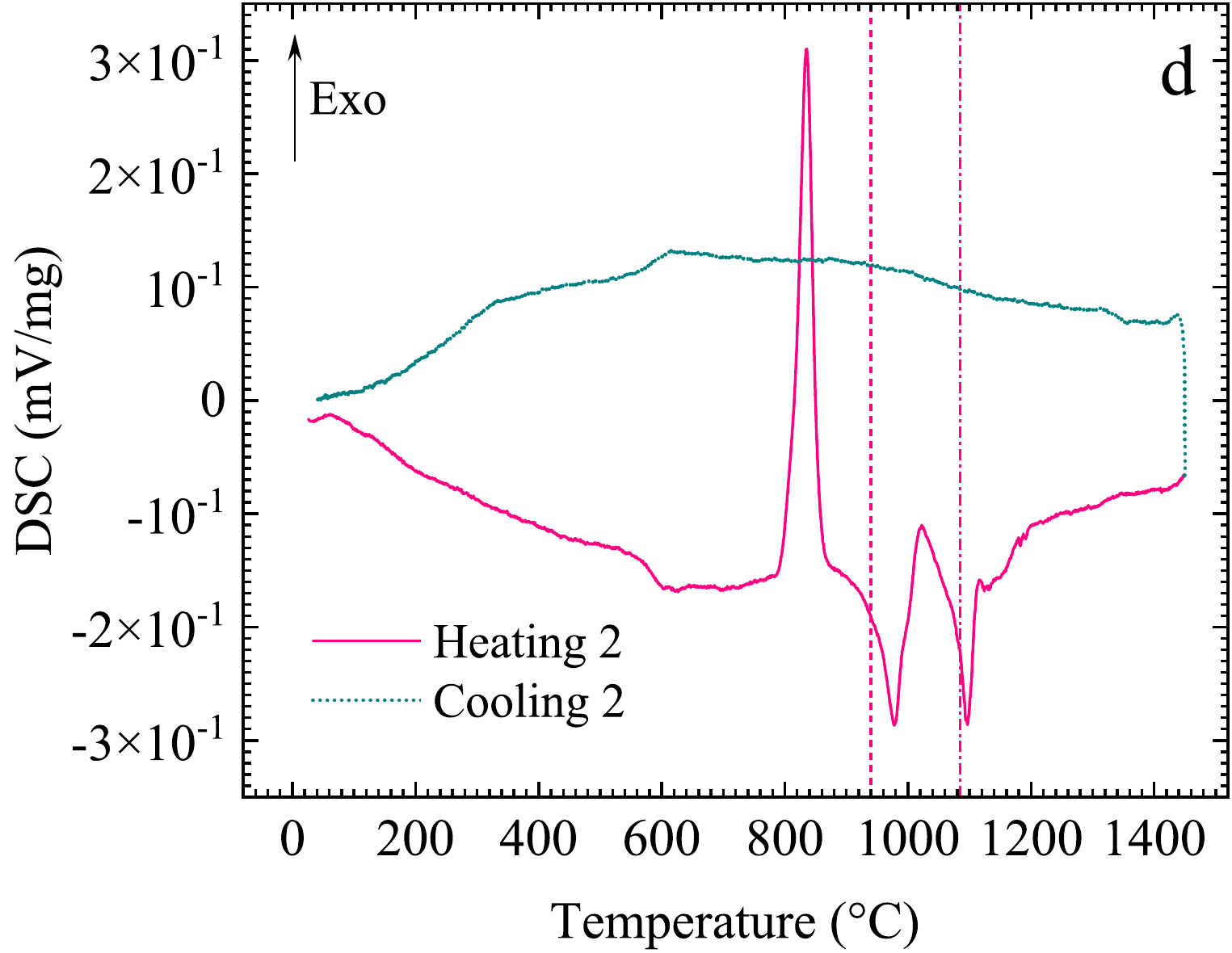}%
						\end{tabular}
		\caption{\label{fig:DSC&TCA}	 
			TG (a,\,b) and DSC (c,\,d) curves of cuprorivaite powder, a {\it double-cycle} experiment ($m_{\rm s}=13.0$~mg): 
			cycle 1 (a,\, c)
			and 
			cycle 2 (b,\,d);
			when {\it heating in cycle~1}, 
			the TG curve shows two steps (a): 
			a strong one with the change of mass of $-0.62$\,\% 
			and a weaker one with the change of mass of $-0.084$\,\% 
			the total change of mass was $-2.0$\,\% and the residual mass was $97.14$\,\%,
			the DSC curve demonstrates two endothermic peaks (c): 
			a strong one at $T=1064.4$\,{\textcelsius} with the onset at $T=1021.3$\,{\textcelsius}, 
			and a weak one at $T=1134.9$\,{\textcelsius}; 
			during {\it cooling in cycle 1} (a,\, c), no features are registered on the curves;
			when {\it heating in cycle 2}, 
			the TG curve shows two steps (b) 
			with the change of mass of $-0.15$\,\% and $-0.11$\,
			the total change of mass was $-1.06$\,\% and the residual mass was $97.06$\,\%, 
			the DSC curve demonstrates an exothermic peak and two endothermic ones (d): 
			a strong exothermic peak at $T=835.4$\,{\textcelsius} with the onset at $T=812.1$\,{\textcelsius}, 
			and endothermic ones at $T=977.5$ and 1096.8\,{\textcelsius} with the onsets at $T\sim940$ and $T=1084.8$\,{\textcelsius}, respectively; 
			when {\it cooling in cycle 2} (b,\, d), as in cycle 1, no features are observed on the curves. 
	}
	\end{minipage}
\end{figure}

The double-cycle experiment allowed us to reveal numerous features appearing on the curves of thermal analysis as a result of repeated heating and cooling, which are signatures of successive chemical and physical transformations of crystalline CaCuSi$_4$O$_{10}$ into a set of phases that differ in both chemical composition and physical state.
Fig.~\ref{fig:DSC&TCA} shows TG and DSC curves of cuprorivaite powder obtained during the double-cycle experiment.

The TG curve recorded during sample heating in the first cycle demonstrates a single pronounced step with a change of mass of $-0.62$\,\% 
starting at a temperature of approximately 1020\,{\textcelsius} and a much weaker one with a mass change of $-0.084$\,\% 
with an onset at about 1133\,{\textcelsius};
the overall mass change during sample heating was approximately $-2.0$\,\% and the residual mass of the sample at the maximum temperature was $97.14$\,\%
(Fig.~\ref{fig:DSC&TCA}\,a).

On the DSC curve (Fig.~\ref{fig:DSC&TCA}\,c), a strong sharp endothermic peak with a maximum at $T=1064.4$\,{\textcelsius} and the onset at $T=1021.3$\,{\textcelsius}, which corresponds to the first step of the TG curve, is observed.
Another weak endothermic peak at $T=1134.9$\,{\textcelsius}, which corresponds to the second step on the TG curve, is also recognizable on the DSC curve.

No fine features have been registered on the curves when cooling the sample to the room temperature in the first cycle; 
nevertheless a broad step with a total mass increase of about 1.0\,\% is visible on the TG curve in the temperature range from $\sim1300$ to $\sim700$\,{\textcelsius}.

Reheating to 1450\,{\textcelsius} (cycle 2), in contrast to heating in cycle 1, revealed a narrow exothermic DSC peak at $T=835.4$\,{\textcelsius} (the onset at 812.1\,{\textcelsius}) without change of mass, an endothermic peak at $T=977.5$\,{\textcelsius} (the onset at $\sim940$\,{\textcelsius}) with 0.15\,\% mass loss and an endothermic peak at 1096.8\,{\textcelsius} (the onset at $\sim1084.8$\,{\textcelsius}) with 0.11\,\% mass loss (Fig.~\ref{fig:DSC&TCA}\,b,\,d)---note the steps starting at $\sim 940$ and $\sim 1085$\,{\textcelsius} on the TG curve.
A weak endothermic peak at $T=1134.9$\,{\textcelsius} is also registered during the second heating.
The total loss of mass during heating in cycle 2 was 1.06\,\% and the residual mass was $97.06$\,\%.

As in the first cycle, fine features have not been registered on the curves when cooling the sample to the room temperature in the second cycle (Fig.~\ref{fig:DSC&TCA}\,b,\,d);
however, on the TG curve in the range from $\sim1350$ to $\sim650$\,{\textcelsius}, there is a broad step with an overall increase in mass of 0.94\,\%.

\subsubsection{Single-cycle experiment} \label{subsubsec:single}

\begin{figure}[t]
	\begin{minipage}[l]{1\textwidth}{\vspace{-0.0cm}}{\hspace{-0cm}}
		\begin{tabular}{rr}
			\includegraphics[height=4cm]{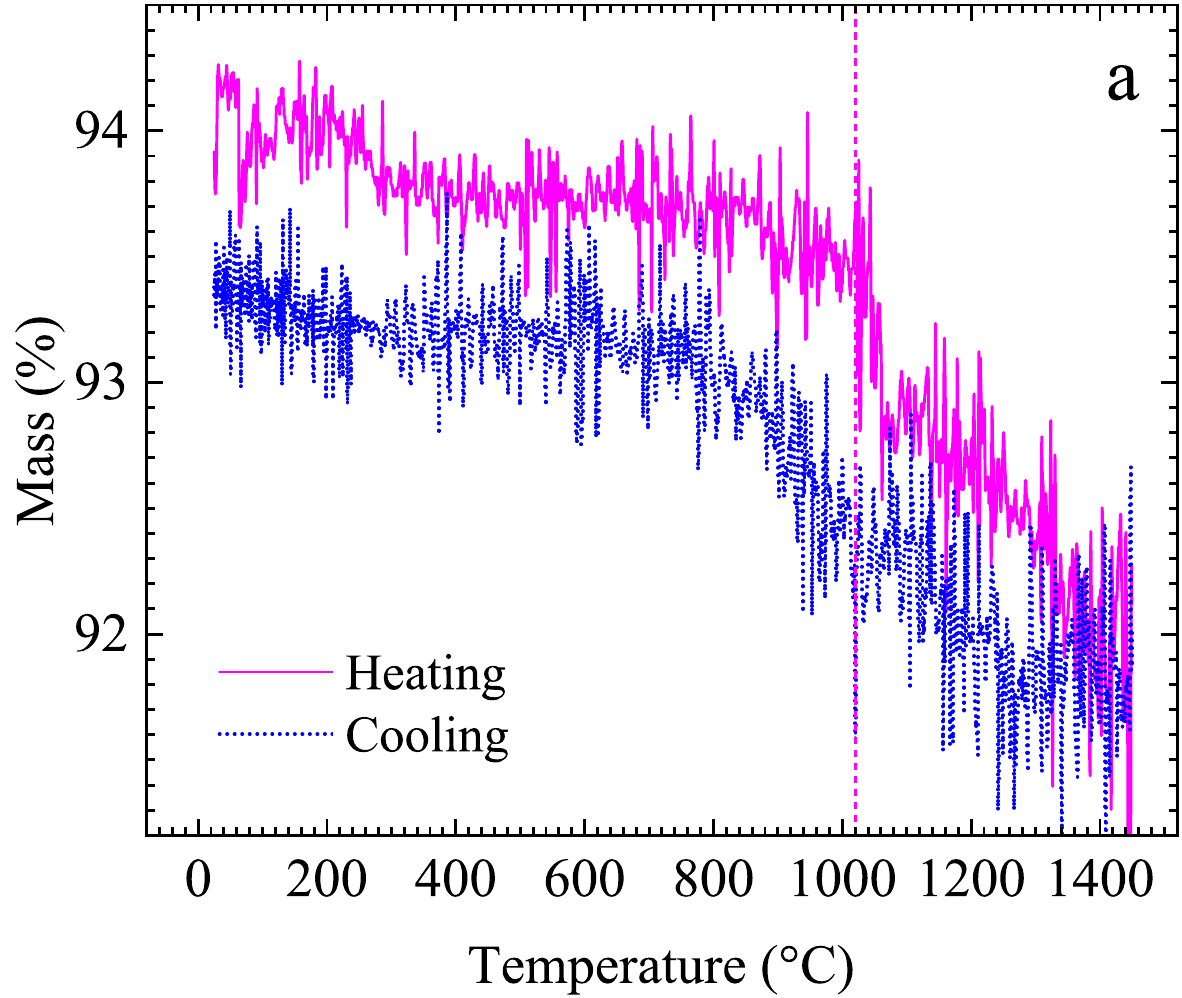} &
			\includegraphics[height=4cm]{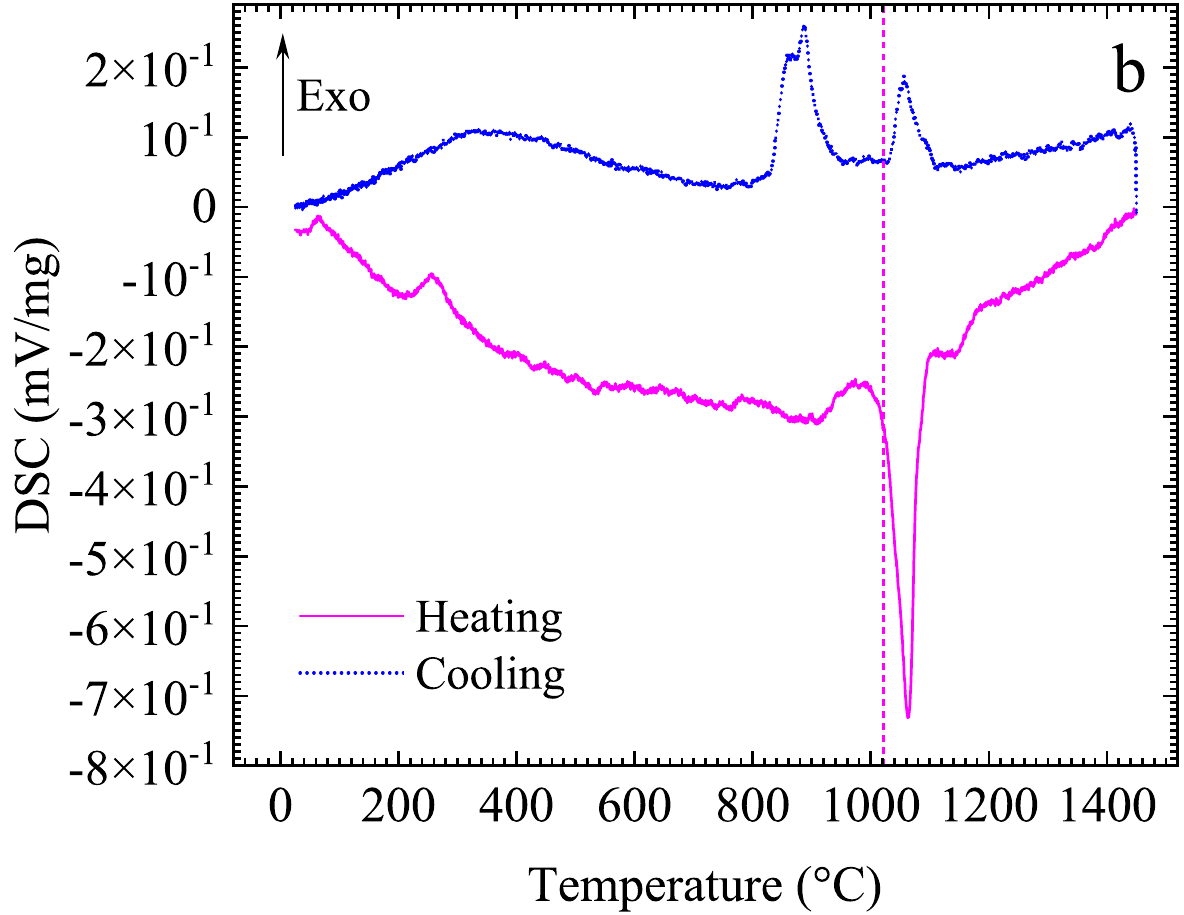}\\
		\end{tabular}
		\caption{%
			\label{fig:DSC&TCA1}%
			TG (a) and DSC (b) curves of the cuprorivaite powder, a {\it single-cycle} experiment ($m_{\rm s}=4.0$~mg): 
			when {\it heating},
			exothermic peaks at $T=64.7$ and~$256.2$\,{\textcelsius} are observed in addition to the 
			endothermic ones at $T=1064.4$ and 1134.9\,{\textcelsius} on the DSC curve;
			the TG curve, although quite noisy, is alike to that obtained during the first heating in the double-cycle experiment;
			when {\it cooling}, in contrast to the double-cycle experiment,
			four extra exothermic peaks with the maxima at $T=1096.8, 1056.0$, 886.3 and 861.5\,{\textcelsius} are revealed on the DSC curve;			
			no mass change corresponds to these peaks. 
					}
	\end{minipage}
\end{figure}

In the single-cycle experiment, the sample mass was considerably less than in the double-cycle one, which enabled recording some extra features on the curves and clarifying the chemical processes occurring during heating and cooling of the sample Fig.~\ref{fig:DSC&TCA1}. 

During heating, in addition to the endothermic peaks at $T=1064.4$ and $T=1134.9$\,{\textcelsius}, which are quite similar to those revealed during the first heating in the double-cycle experiment, two exothermic peaks are observed at $T=64.7$ and 256.2\,{\textcelsius} on the DSC curve.
Note that very weak exothermic features could be detected at these temperatures on the DSC curve during the first heating in the double-cycle experiment.
The TG curve does not show any mass change that could be associated with these low-temperature transformations.
In general, the TG curve, although quite noisy, is similar to that obtained during the first heating in the double-cycle experiment.
An overall change of mass was about $-2.1(3)$\,\%.

When cooling, in contrast to the double-cycle experiment, four extra exothermic peaks with the maxima at $T=1096.8$, 1056.0, 886.3 and 861.5\,{\textcelsius} appear on the DSC curve;			
no mass change corresponding to these peaks was resolved, possibly due to noise in the TG curve. 
Overall change of mass was approximately $+1.5(1)$\,\%.

The data of the DSC and TG analyses are gathered in Table~\ref{tab:Peaks}.

\subsection{Phase analysis} \label{sec:XRD}

\begin{figure}[t]
\includegraphics[width=0.32\linewidth]{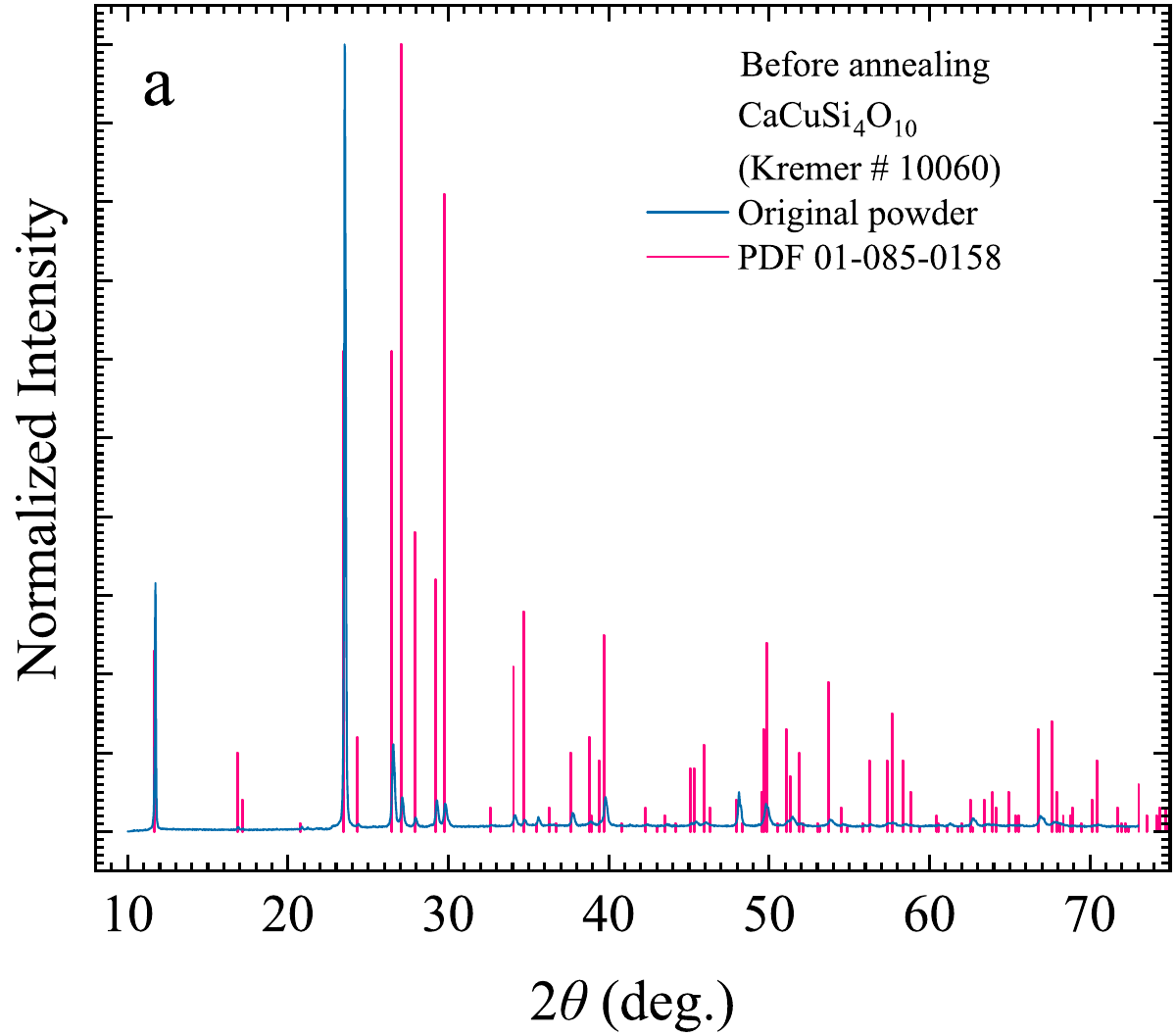}
\includegraphics[width=0.32\linewidth]{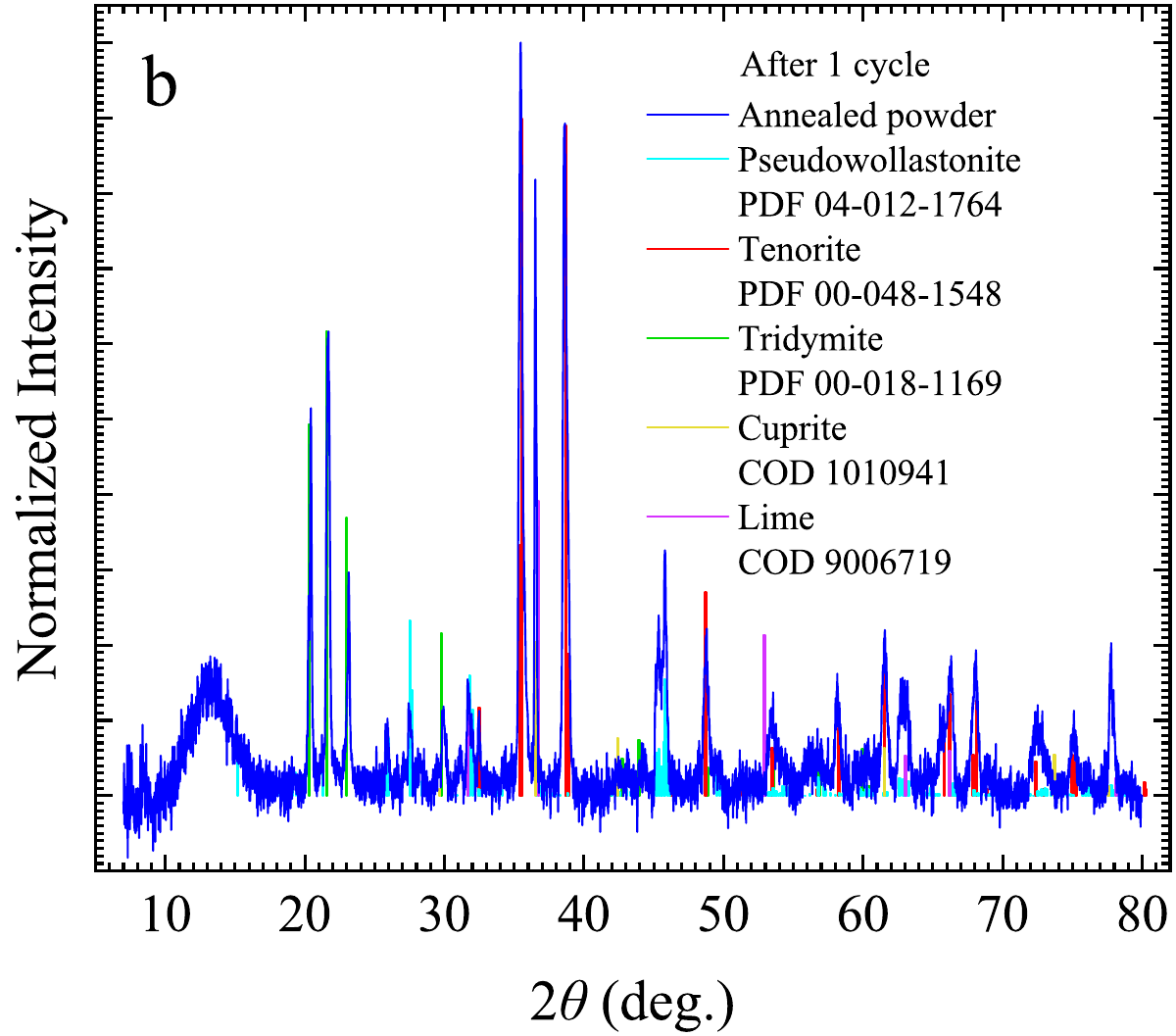}
\includegraphics[width=0.32\linewidth]{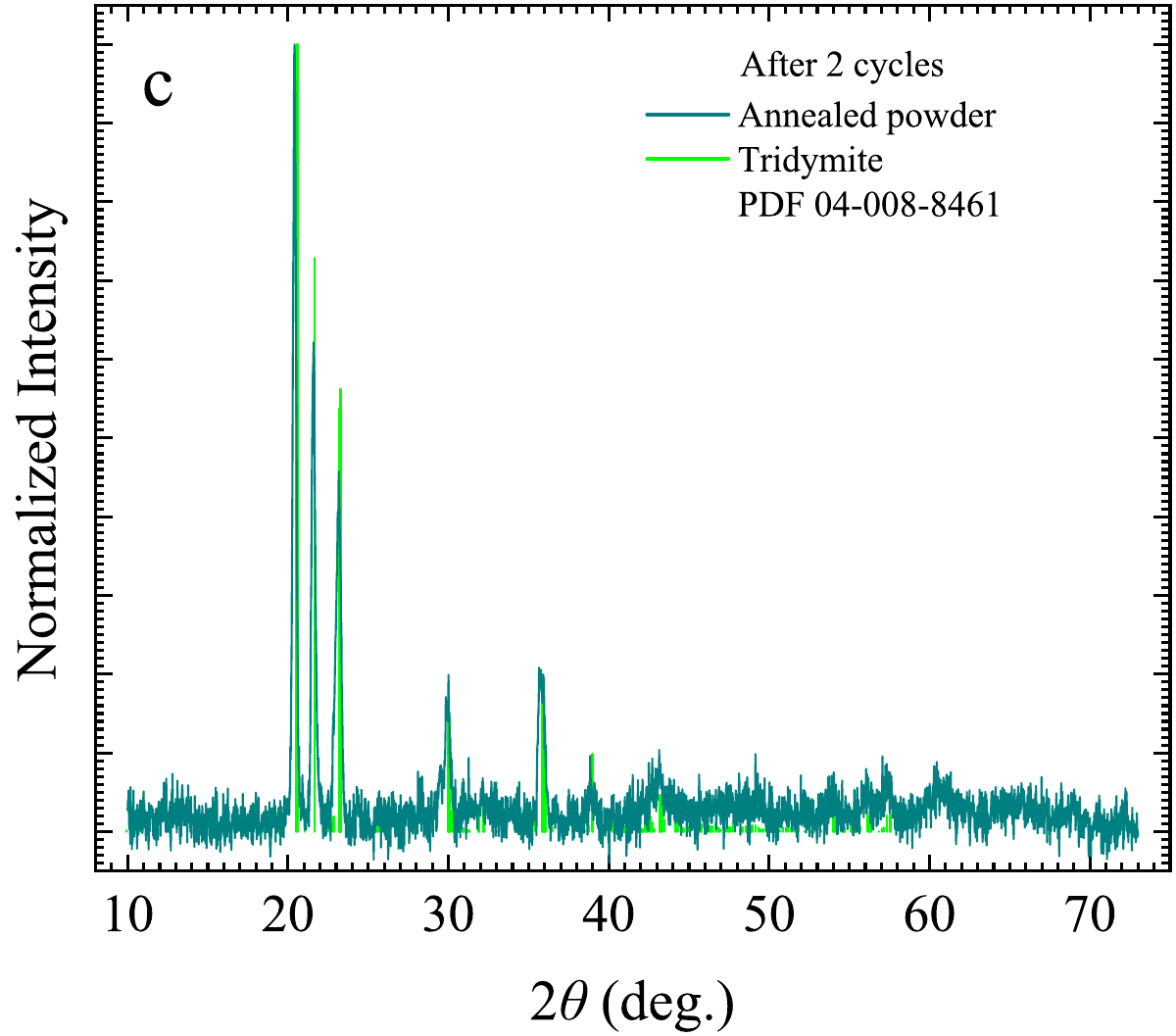}
	\caption{X-ray powder patterns (Cu\,K$_{\alpha}$) of the sample
		(a) before the thermal analysis (original powder), (b) after the first cycle (a single-cycle experiment) and (c) after the second cycle (a double-cycle experiment).
		\label{fig:XPA}	 
	}
\end{figure}

X-ray phase analysis of the resultant product was performed before and after each experiment.
Fig.~\ref{fig:XPA} shows the obtained XRD patterns; 
the identified phases are shown in the graphs;
more detailed data, including the characteristics of the crystalline phases as well as the percentage content of the detected substances estimated using the reference intensity ratio method (RIR), are given in Table~\ref{tab:Phases}.

The XRD pattern of the original sample (Fig.~\ref{fig:XPA}\,a) showed only the cuprorivaite phase (PDF 01-085-0158, \cite{Bensch_cuprorivaite}).

The diffraction pattern obtained after sample cooling in the single-cycle experiment has demonstrated a much more diverse picture of phase composition (Fig.~\ref{fig:XPA}\,b).
The crucible contained the following main crystalline phases: 
synthetic pseudowollastonite (CaSiO$_3$, PDF 04-012-1764, \cite{pseudowollastonite}) in an amount of 32.6\,wt.\%; 
synthetic tenorite (CuO, PDF 00-048-1548,	\cite{Tenorite_00-048-1548}) in an amount of 32.2\,wt.\%;
and
high-temperature synthetic hexagonal tridymite (SiO$_2$, PDF 00-018-1169, \cite{Sato_tridymite, Kihara_tridymite}) in an amount of 28.3\,wt.\%.
Two additional minor phases have also been identified in the crucible:
cuprite (Cu$_2$O, COD~1010941, \cite{Niggli_Cuprite}) in an amount of 5.6\,wt.\%
and
lime (CaO, COD~9006719, \cite{Fiquet_Lime}) in an amount of about 1.0\,wt.\%.

Finally, Fig.~\ref{fig:XPA}\,c shows that the only crystalline phase present in the crucible after the second cooling of the sample in the double-cycle process was low-temperature monoclinic tridymite (SiO$_2$, PDF 04-008-8461, \cite{Dollase_tridymite}). 

\subsection{Analysis using polarization microscopy} \label{subsec:polar_micr}

\subsubsection{After the first cycle of annealing} \label{subsubsec:polar_cycle_1}

\begin{figure}[t]
	\includegraphics[width=0.4\linewidth]{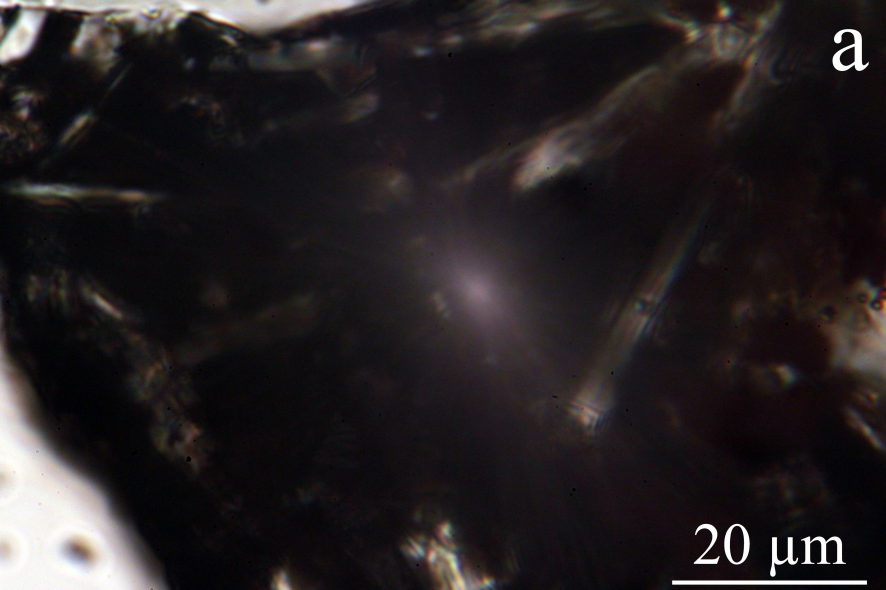}
	\includegraphics[width=0.4\linewidth]{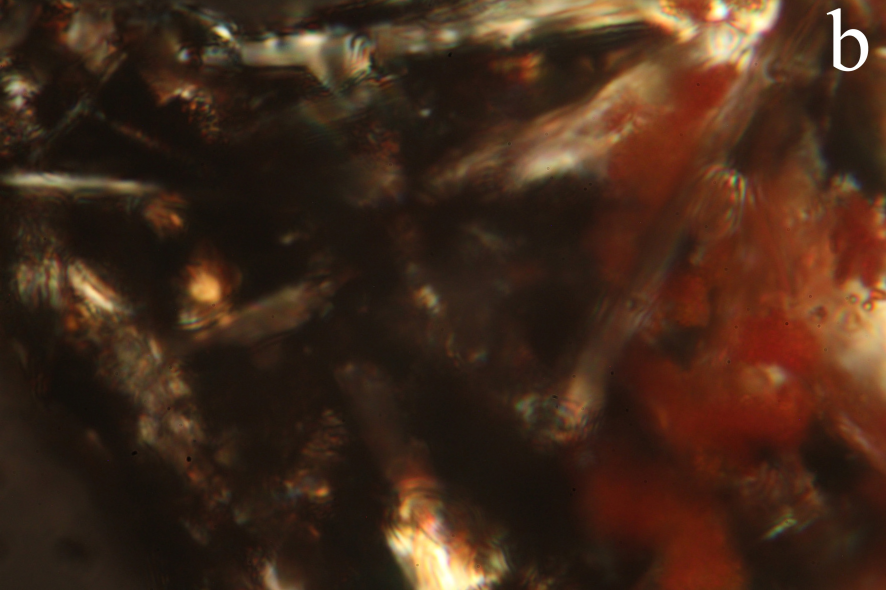}\\
	\includegraphics[width=0.4\linewidth]{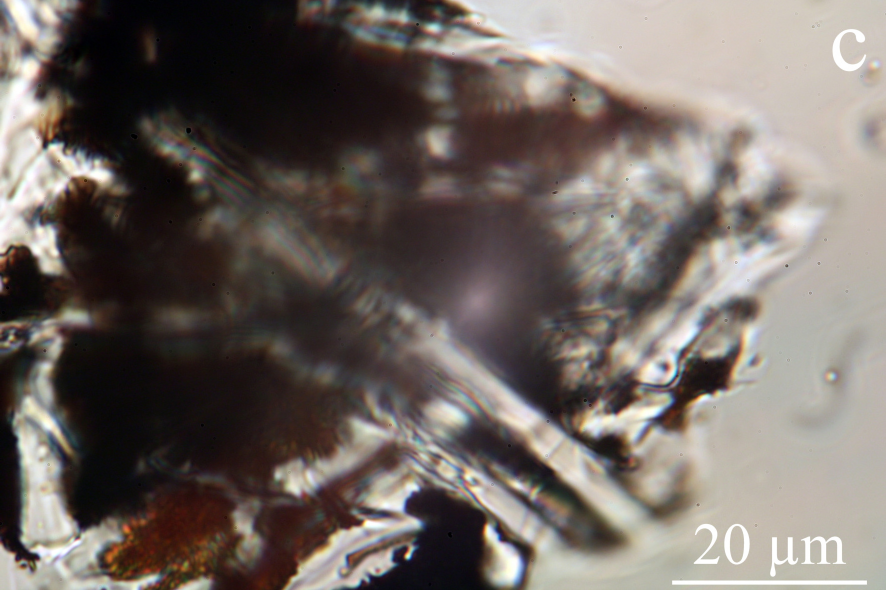}
	\includegraphics[width=0.4\linewidth]{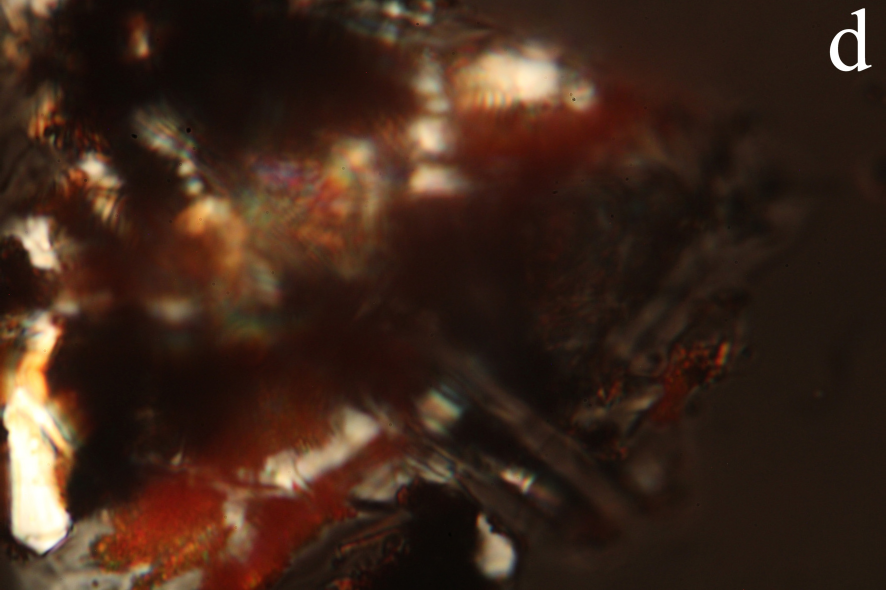}\\
	\includegraphics[width=0.4\linewidth]{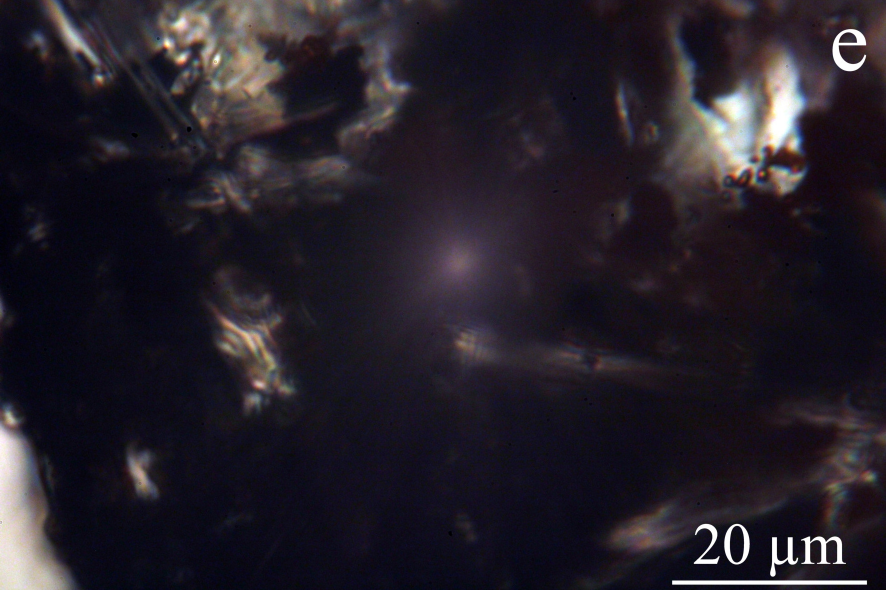}
	\includegraphics[width=0.4\linewidth]{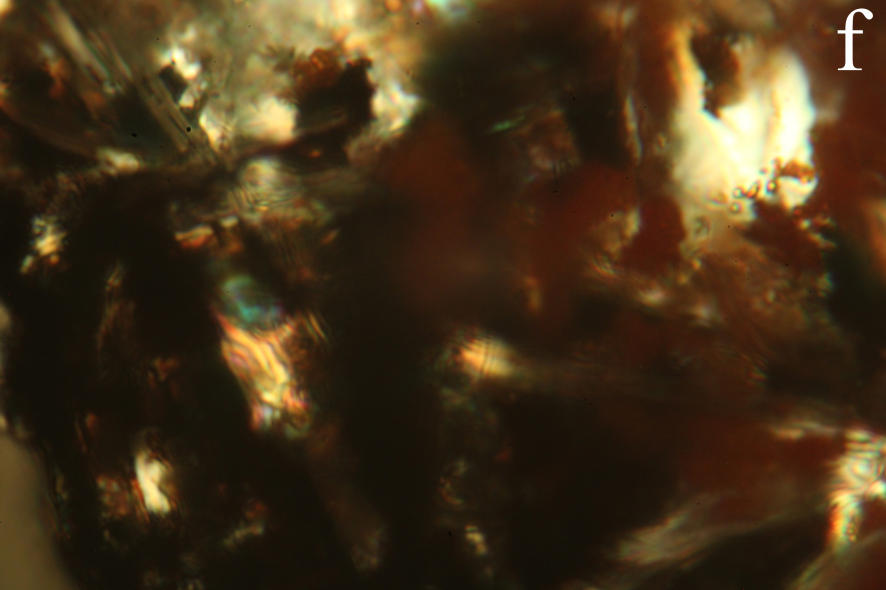}\\
	\includegraphics[width=0.4\linewidth]{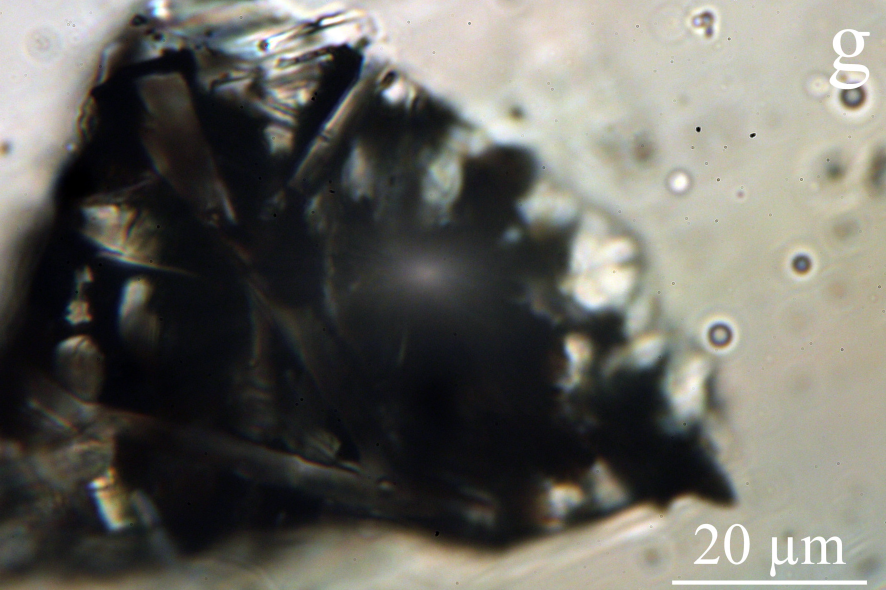}
		\includegraphics[width=0.4\linewidth]{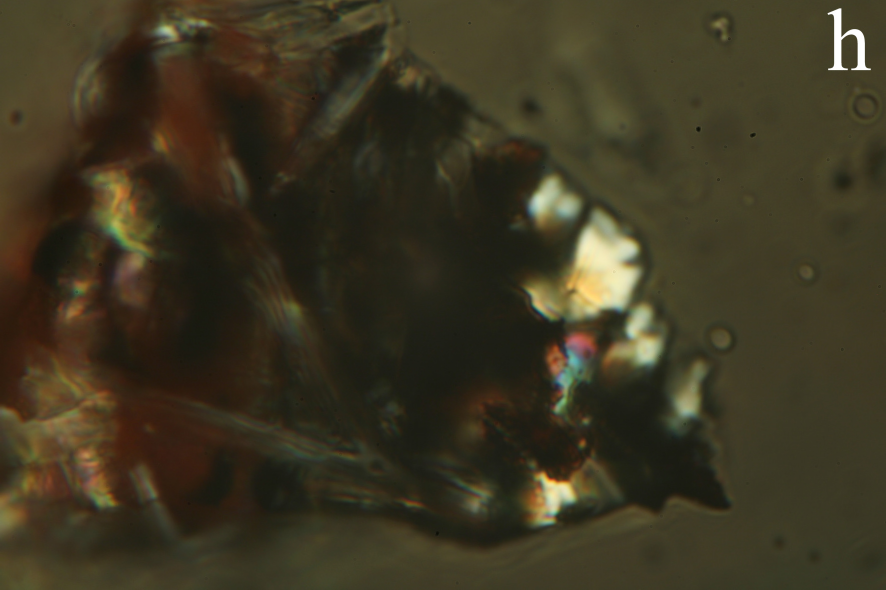}
	\caption{%
		Complementary pairs of images of sample fragments after the single-cycle experiment, obtained using polarization light microscopy: parallel (a,\,c,\,e,\,g) and crossed (b,\,d,\,f,\,h) polars.
		\label{fig:PM_02-18_a}	 
	}
\end{figure}

Fig.\ref{fig:PM_02-18_a} represents complementary pairs of PM images of sample fragments after the single-cycle annealing.
In each pair, the images on the left are micrographs obtained with parallel polarizers, and the those on the right are micrographs obtained with crossed polarizers.
Images taken with parallel polarizers show that the annealing process resulted in a multiphase composite, with some phases completely absorbing light, while others being partially or almost completely transparent.
It can be seen that the samples contain both acicular crystals and areas that look like amorphous formations.

Micrographs obtained with crossed polarizers are significantly more informative.
They show the same acicular crystals as the images taken with parallel polarizers.
These crystals contain numerous twins and have no specific coloration---they typically appear light gray and show virtually no color change along their entire length; 
they do not exhibit interference.
In addition, the images show black areas that appear almost white or light gray in photographs taken with parallel polarizers, consistent with a transparent substance that is not birefringent.
Other features noticeable in these images are bright areas that change color from white to yellowish. 
Birefringence patterns are observed under crossed polars in these areas.
Some of them are visible as transparent platy particles or simply bright white domains under parallel polars.
Finally, these images particularly highlight areas of dendritic structures that are bright red or black.
These structures are not visible in photographs taken under parallel polars because they strongly absorb light, and the areas of the samples containing them appear black or dark gray.
However, the red domains sometimes do not show a dendritic structure, but on the contrary, look granular, both under parallel and crossed polarizers (Fig.\ref{fig:PM_02-18_a}\,c,\,d), while the black areas often also do not look dendritic but rather resemble stripes, alternating with white or gray bands, when photographed under both parallel and crossed polars (Fig.\ref{fig:PM_02-18_a}\,g,\,h).

\begin{figure}[t]
	\includegraphics[width=0.2\linewidth]{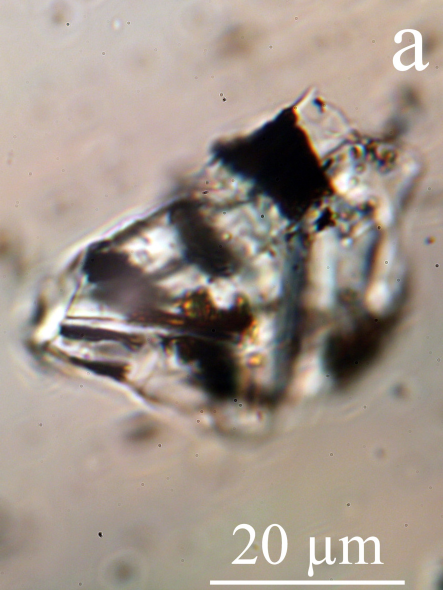}%
	\includegraphics[width=0.2\linewidth]{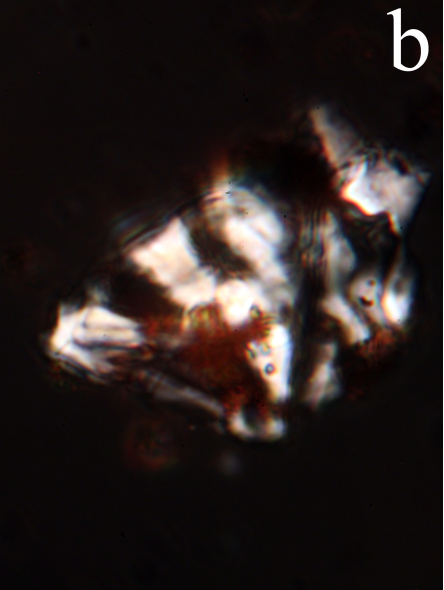}
	\includegraphics[width=0.2\linewidth]{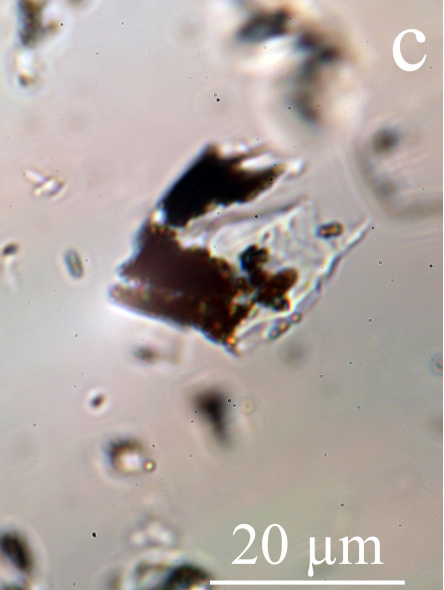}%
	\includegraphics[width=0.2\linewidth]{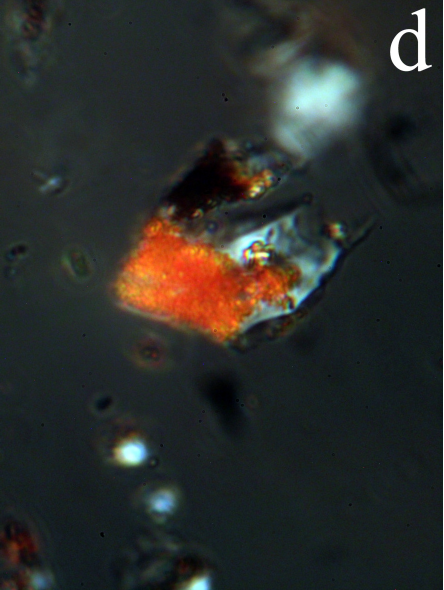}\\
	\includegraphics[width=0.4\linewidth]{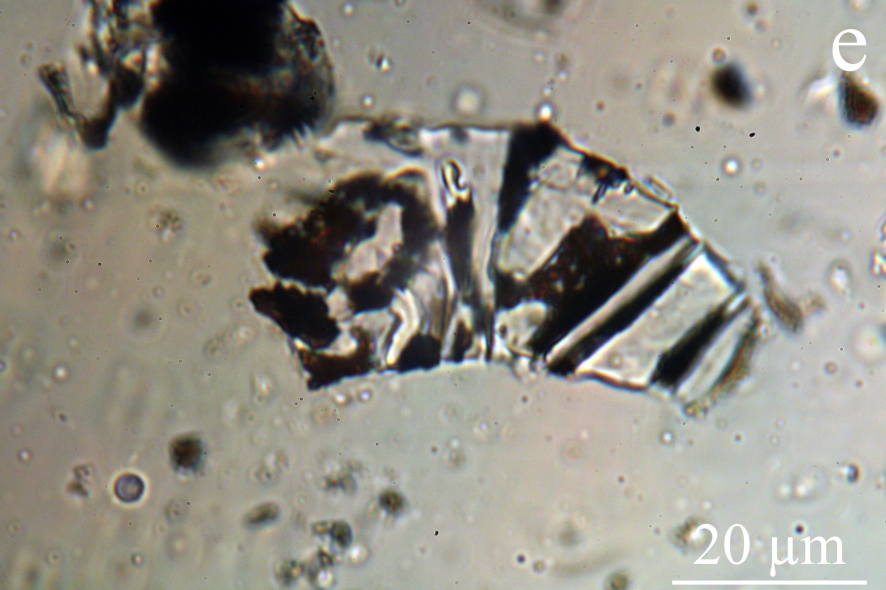}
	\includegraphics[width=0.4\linewidth]{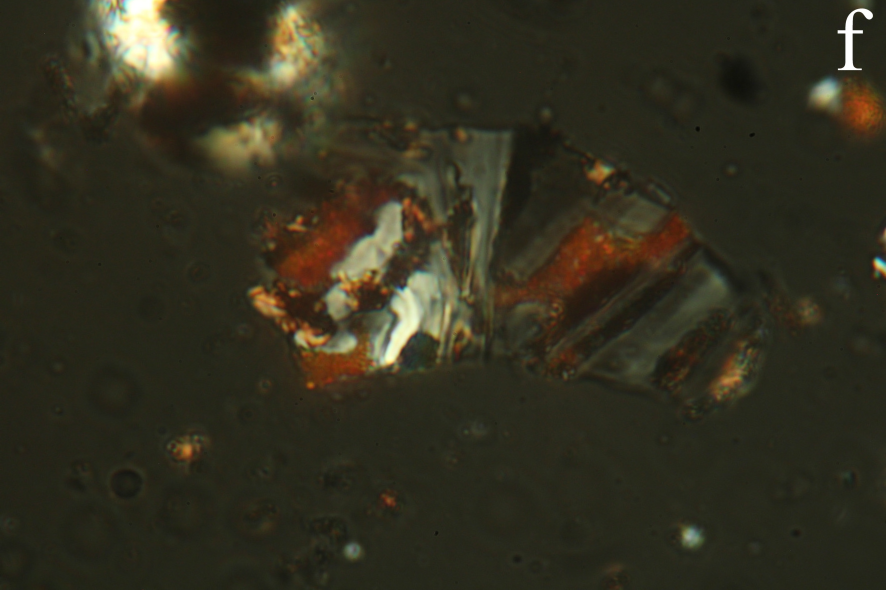}\\
	\includegraphics[width=0.4\linewidth]{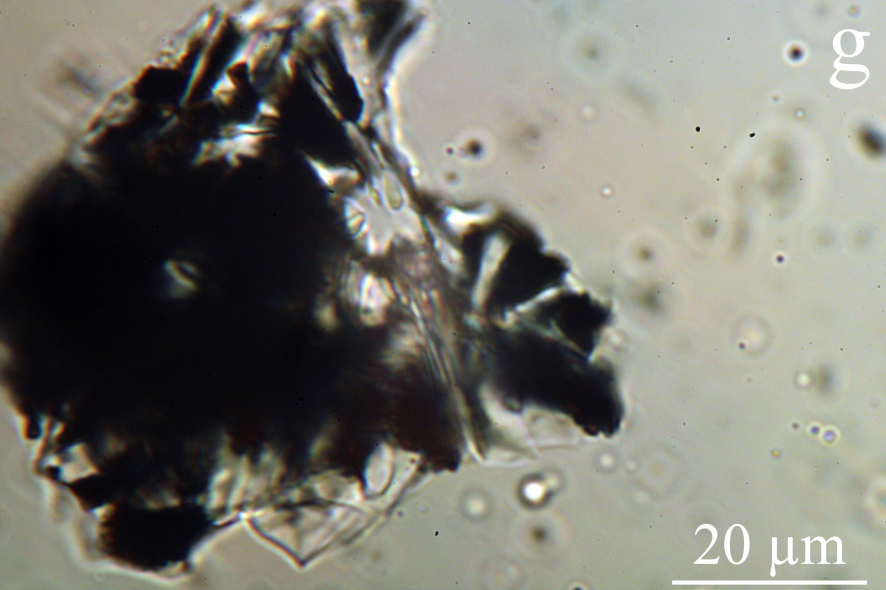}
	\includegraphics[width=0.4\linewidth]{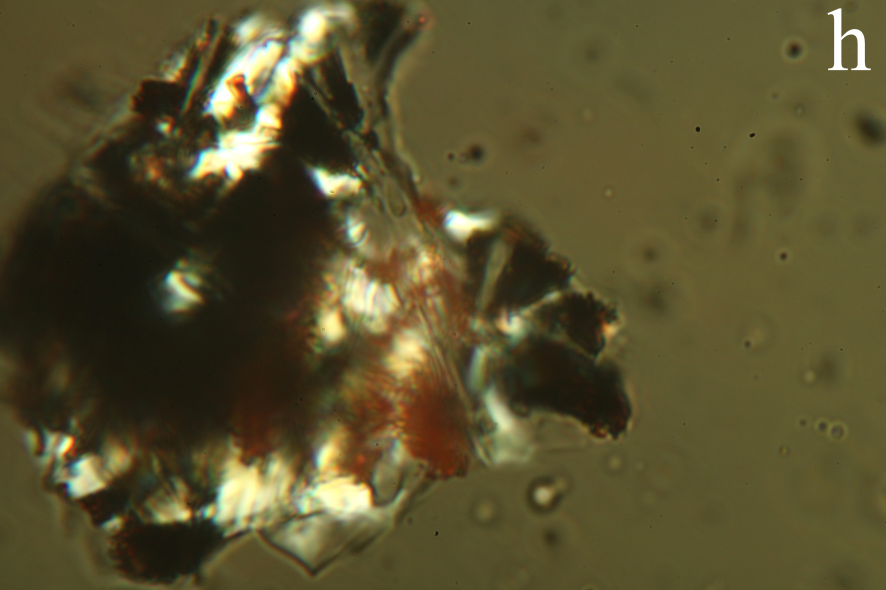}
	\caption{%
		Complementary pairs of PM images of small pieces of the substance after the single-cycle experiment obtained with parallel (a,\,c,\,e,\,g) and crossed (b,\,d,\,f,\,h) polars.
		\label{fig:PM_02-18_b}	 
	}
\end{figure}

In Fig.\ref{fig:PM_02-18_b}, small particles are presented, which also brightly illustrate the above-mentioned features.
Panels a and b show a platy particle, which, when viewed through parallel polars (a), displays transparent bands alternating with black ones, while in crossed polars (b), they appear white and black; 
dark red grainy areas are also observed in panel (b).
A platy particle with black and transparent areas (c) demonstrate a domain consisting of red grains instead of the black domain as well as blue domain instead of the transparent one, when photographed with crossed polars (d); the latter one look blue due to birefringence. 
In the panel e, two platy crystallites are observed showing transparent and black domains, which appear colored black, white, red and bright orange in the panel f, with red and yellow domains being composed of tiny grains.
Finally, a particle that looks consisting of black and transparent regions, when viewed through parallel polars (g), exhibits black, bright white, gray, yellowish and dark red ones with crossed polars (h).

\begin{figure}[t]
	\includegraphics[width=0.4\linewidth]{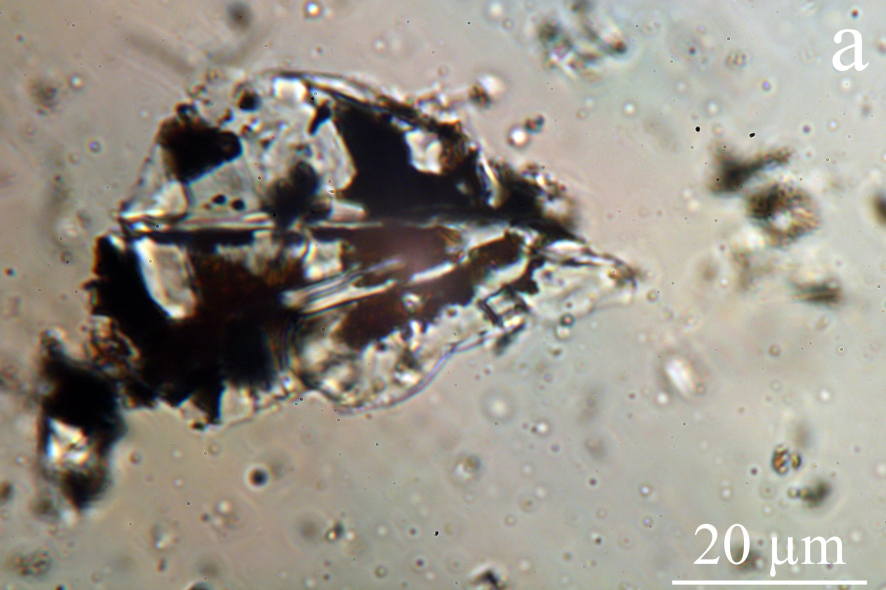}
	\includegraphics[width=0.4\linewidth]{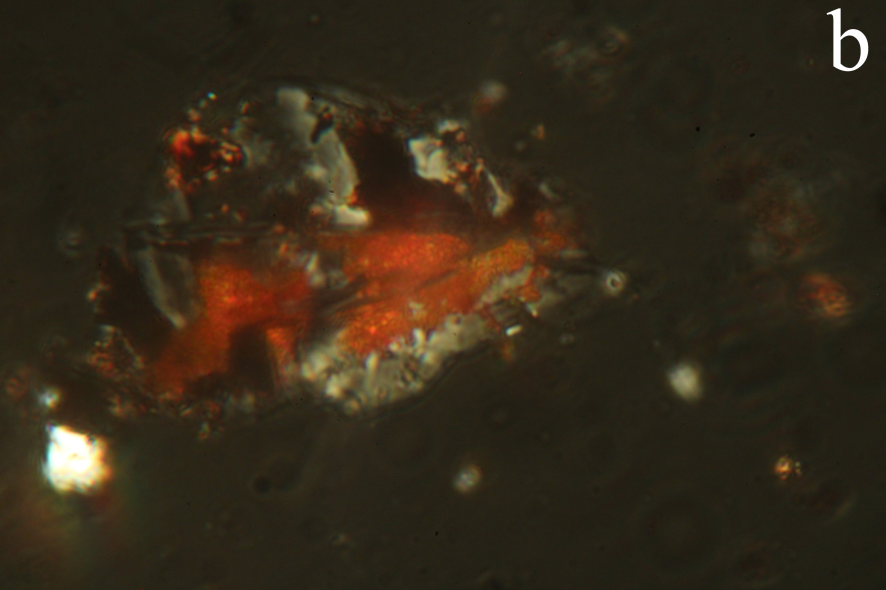}\\
	\includegraphics[width=0.4\linewidth]{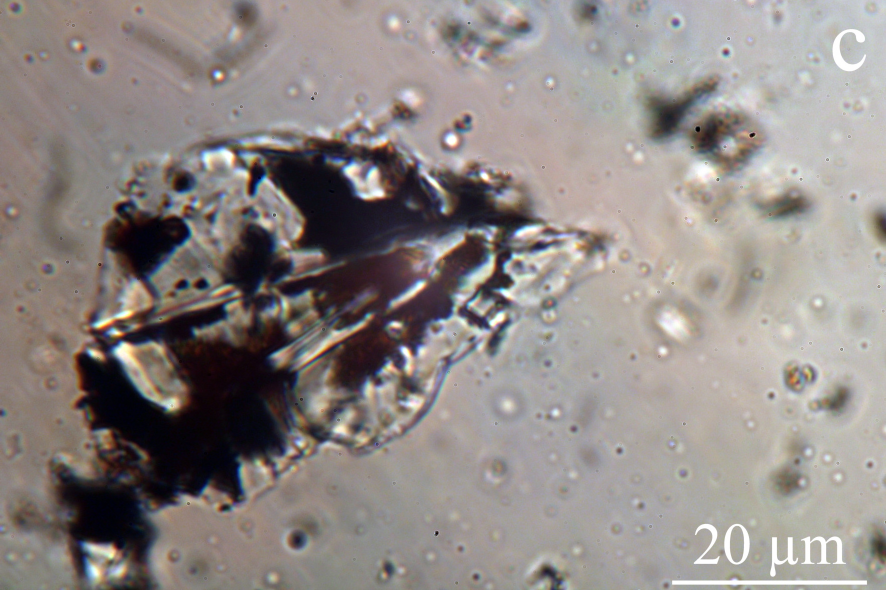}
	\includegraphics[width=0.4\linewidth]{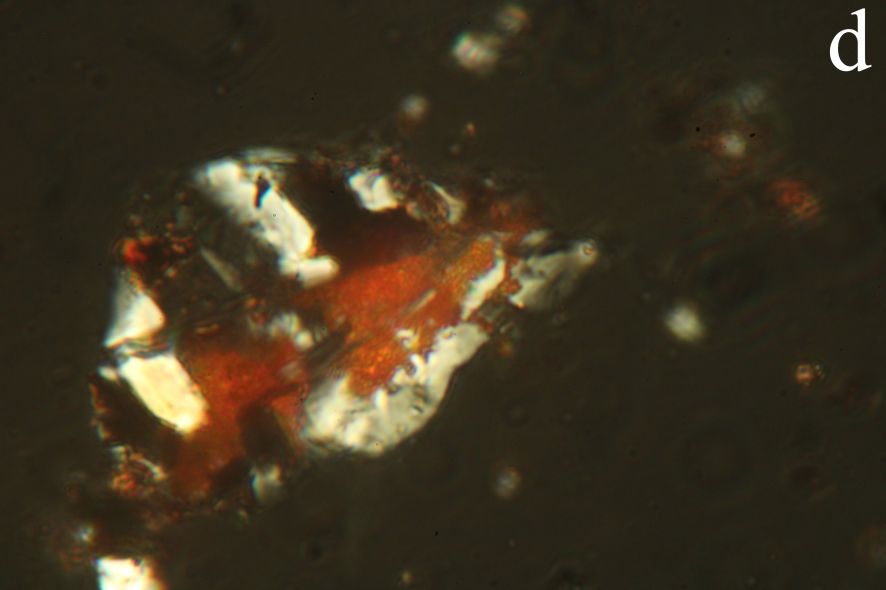}\\
	\includegraphics[width=0.2\linewidth]{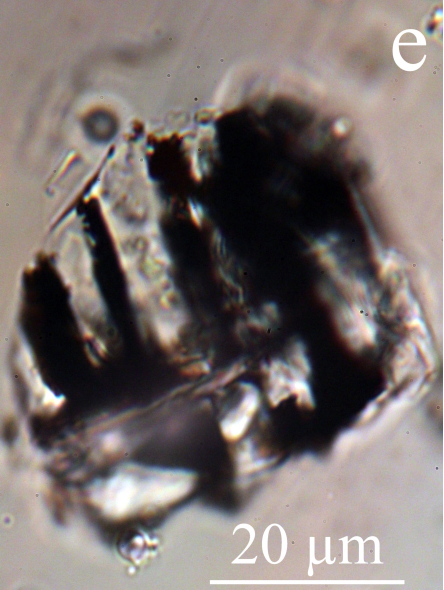}%
	\includegraphics[width=0.2\linewidth]{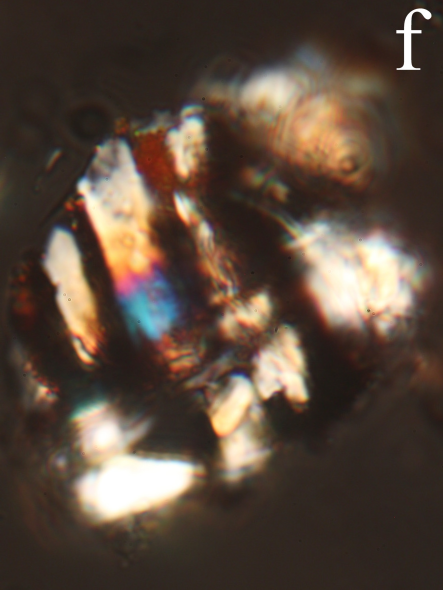}
	\includegraphics[width=0.2\linewidth]{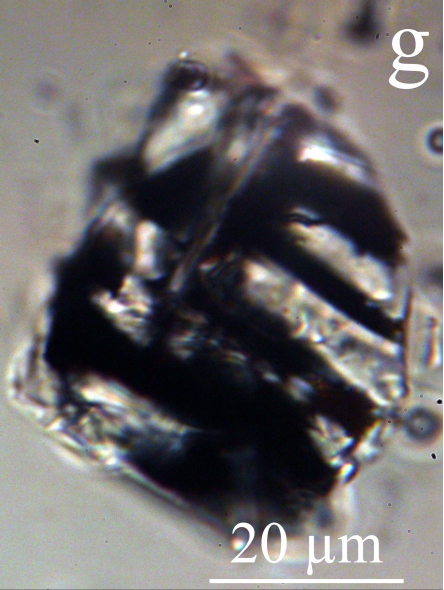}%
	\includegraphics[width=0.2\linewidth]{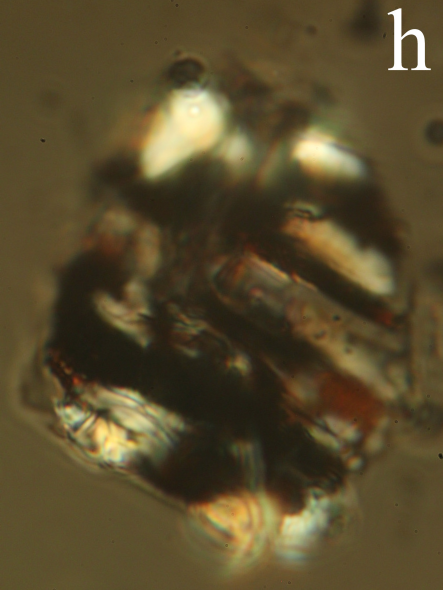}
	\caption{%
		Effect of stage rotation: complementary pairs of PM images of the substance particles after the single-cycle experiment; the micrographs were obtained with parallel (a,\,c,\,e,\,g) and crossed (b,\,d,\,f,\,h) polars.
		\label{fig:PM_02-18_r}	 
	}
\end{figure}

It should also be noted that rotation of the slide led to a change in the color of some areas of the particles when they were observed in crossed polarizers; 
this effect is due to the interference of ordinary and extraordinary rays in birefringent crystals.
Fig.\ref{fig:PM_02-18_r} illustrates the changes in color of different regions of particles due to slide rotation.  
Panels a\,--\,d show changes in shades of gray and white (first-order colors), and in panels e\,--\,h one can see the change in shades of gray and white as well as the disappearance of bright yellow, red and blue (second and higher color orders).

\subsubsection{After the second cycle of annealing} \label{subsubsec:polar_cycle_2}

\begin{figure}[t]
	\includegraphics[width=0.4\linewidth]{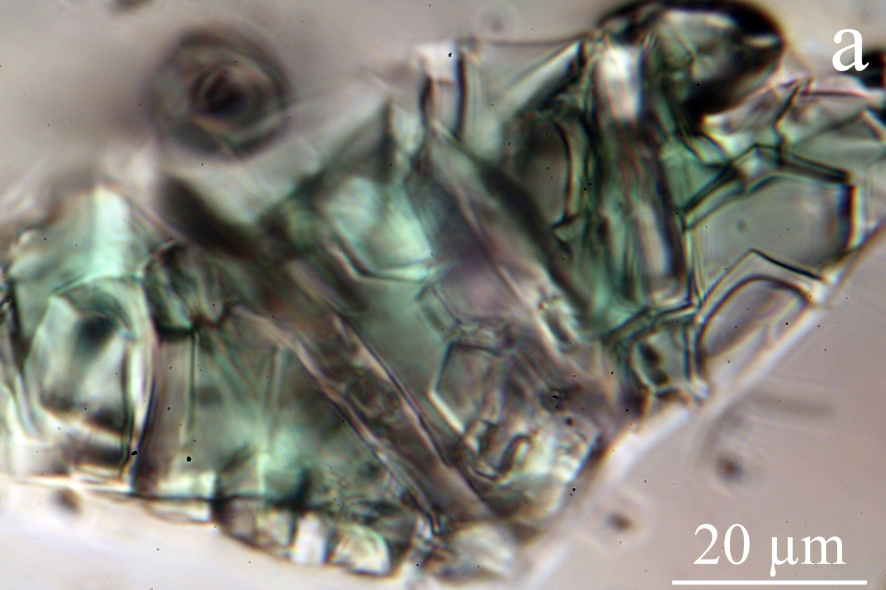}
	\includegraphics[width=0.4\linewidth]{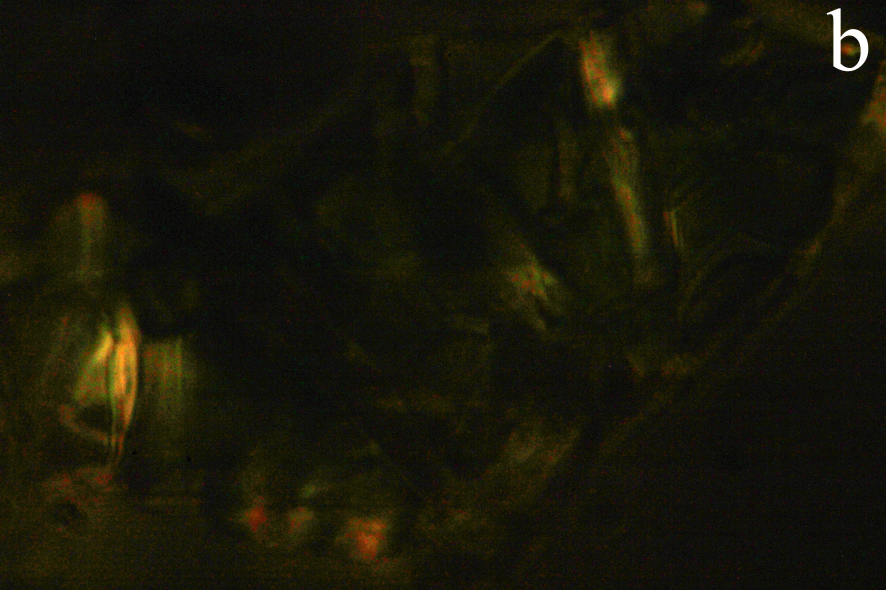}\\
	\includegraphics[width=0.4\linewidth]{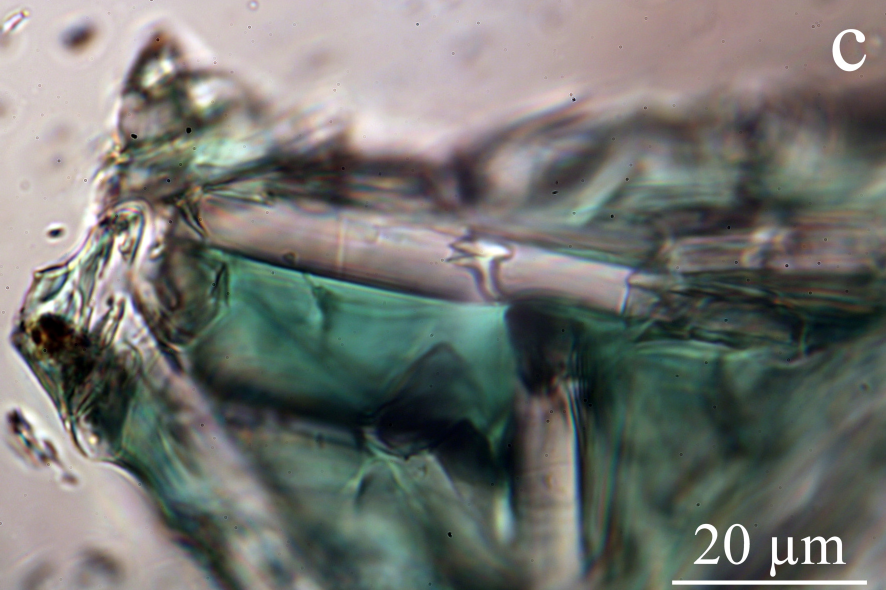}
	\includegraphics[width=0.4\linewidth]{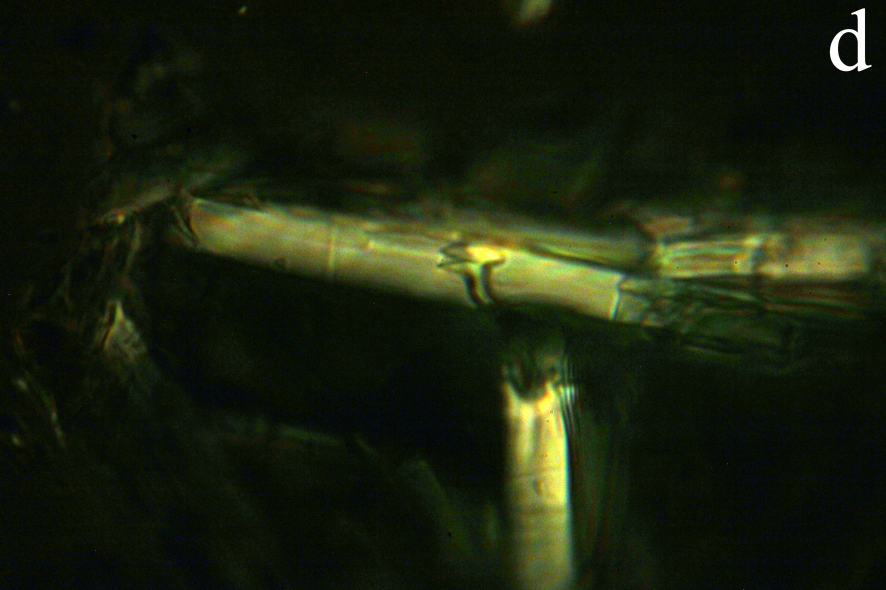}\\
	\includegraphics[width=0.4\linewidth]{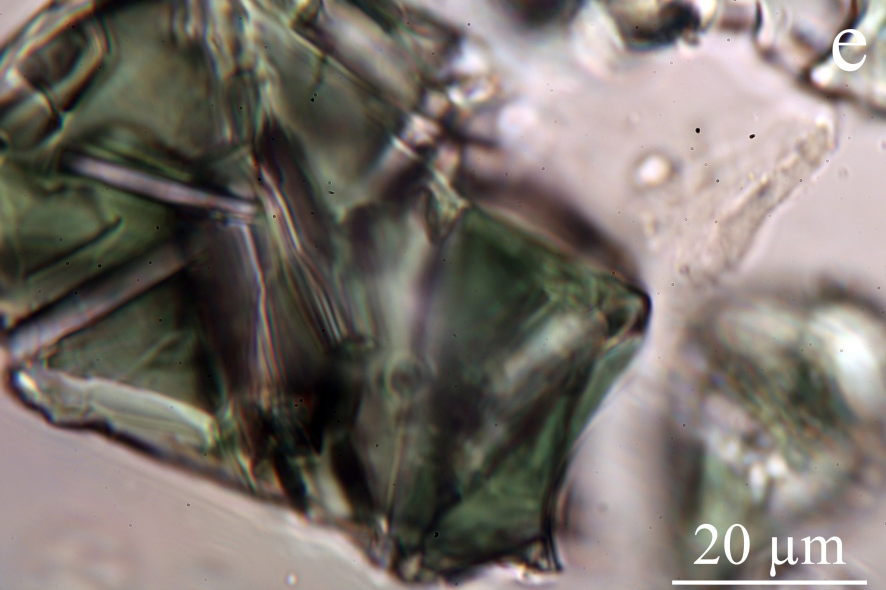}
	\includegraphics[width=0.4\linewidth]{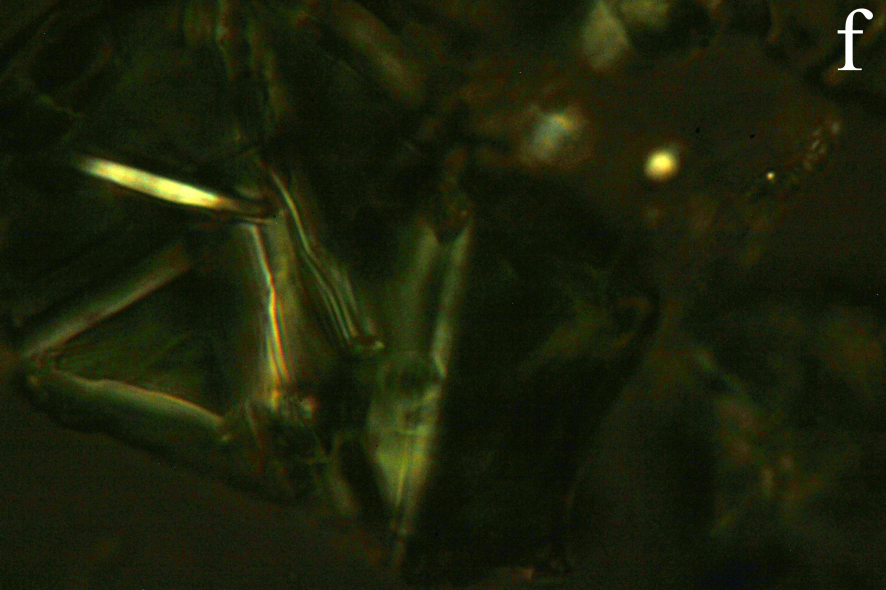}
	\caption{%
		Complementary pairs of PM images of sample fragments after the double-cycle experiment: parallel (a,\,c,\,e) and crossed (b,\,d,\,f) polars.
		\label{fig:PM_07-09_a}	 
	}
\end{figure}

Fig.\ref{fig:PM_07-09_a} presents complementary pairs of PM images of sample fragments after the double-cycle annealing.
As in Figs.~\ref{fig:PM_02-18_a} to~\ref{fig:PM_02-18_r}, in each pair the images on the left are micrographs obtained with parallel polarizers, and the those on the right are micrographs obtained with crossed polarizers.
The images obtained with parallel polars demonstrate a conglomerate of gray acicular crystals cemented with green glass of various tints, whereas only the acicular crystals are observed in the photos obtained with crossed polars, and dark areas are seen instead of the glass.
Sometimes small red or bright orange spots are also visible (panel~b), which appear white in images obtained with parallel polarizers (panel~a).

\begin{figure}[t]
	\includegraphics[width=0.4\linewidth]{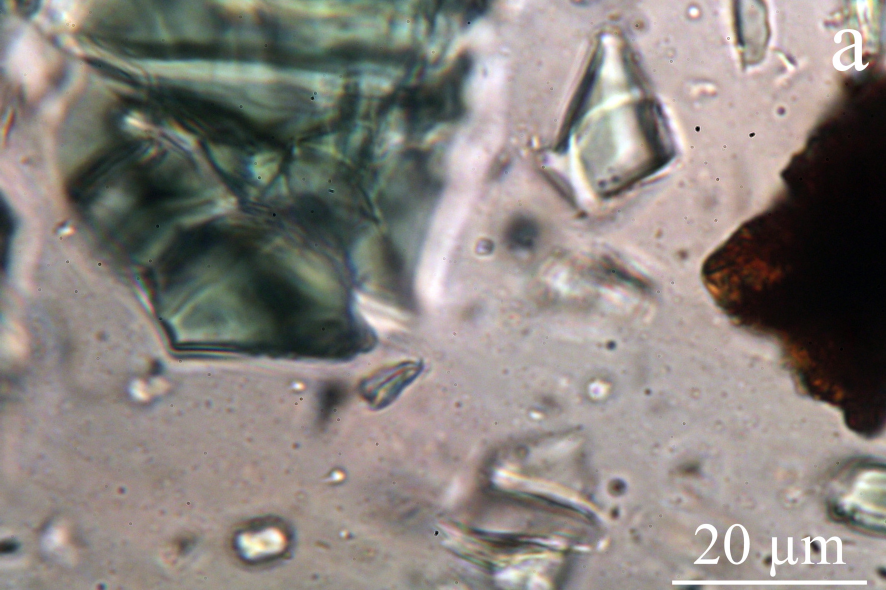}
	\includegraphics[width=0.4\linewidth]{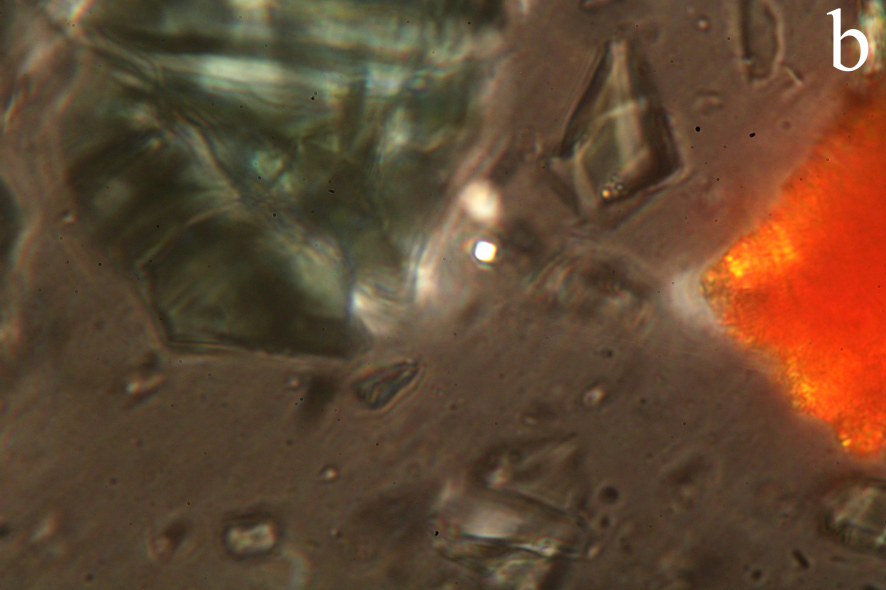}\\
	\includegraphics[width=0.4\linewidth]{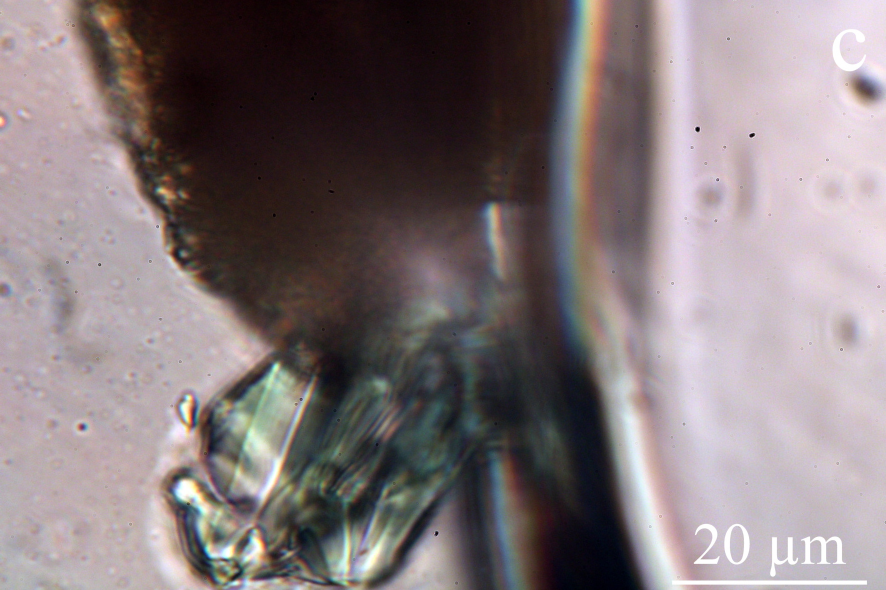}
	\includegraphics[width=0.4\linewidth]{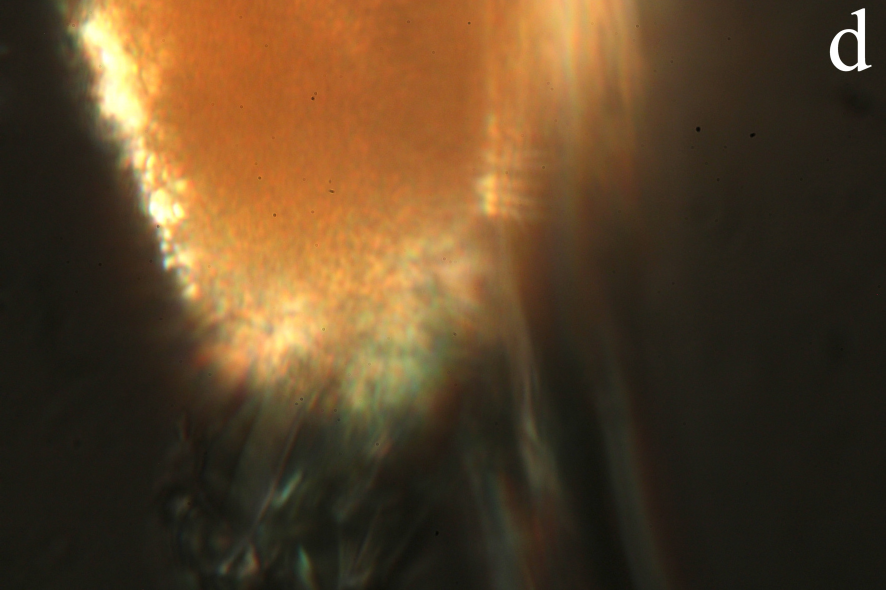}\\
	\includegraphics[width=0.4\linewidth]{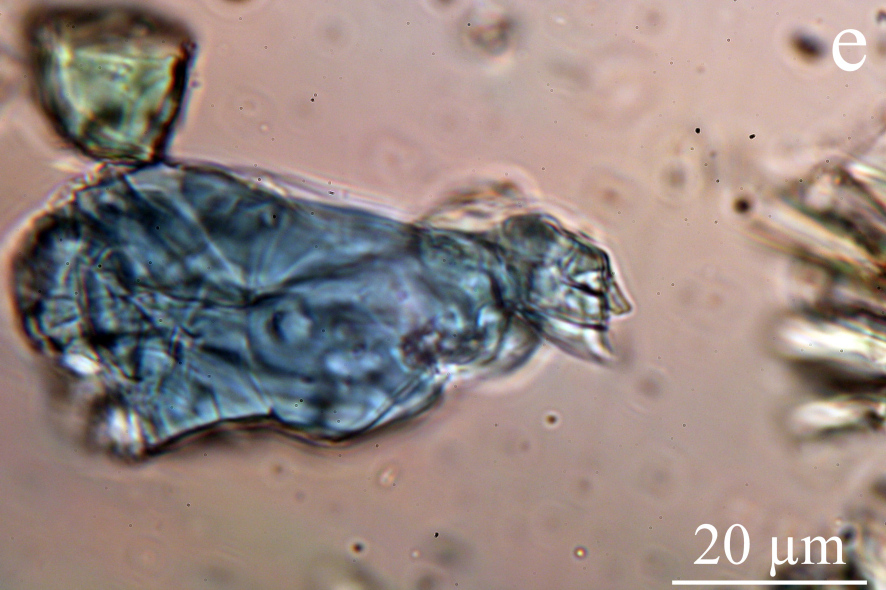}
	\includegraphics[width=0.4\linewidth]{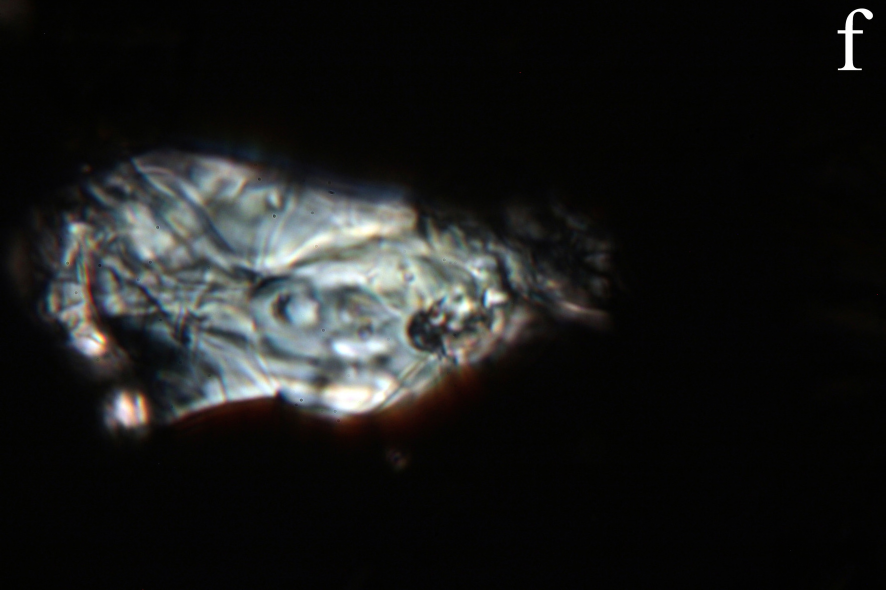}
	\caption{%
		Complementary pairs of polarization microscopy images of sample fragments after the double-cycle experiment: examples of cuprite and cuprorivaite particles; parallel (a,\,c,\,e) and crossed (b,\,d,\,f) polars.
		\label{fig:PM_07-09_b}	 
	}
\end{figure}

However, some small particles of other substances can be found in the preparations in addition to glass with acicular crystals (Fig.\ref{fig:PM_07-09_b}).
Some of them are dark when viewed in parallel polars but bright red or yellow in crossed ones (Fig.\ref{fig:PM_07-09_b}\,a--d).
A few platy particles appear blue in both parallel and crossed polars (Fig.\ref{fig:PM_07-09_b}\,e,\,f), which is characteristic of cuprorivaite (see Fig.\ref{fig:SEM+PM}\,c,\,d).

\subsection{Discussion} \label{subsec:discuss}

\subsubsection{Thermal analysis and X-ray diffraction} \label{subsubsec:TA&XRD}

Let us discuss the nature of the detected DSC peaks and mass changes in the TG curves.
First of all, we should dwell on the endothermic peak at 1064.4\,{\textcelsius} observed in the DSC curves during the first heating of cuprorivaite (Table~\ref{tab:Peaks}).
This peak can be attributed to incongruent melting (decomposition) of cuprorivaite.
It should be noted that early studies stated that Egyptian blue becomes unstable at temperatures above 1050\,{\textcelsius} \cite{EB_technology, tite1987}.
This study enabled determining the exact melting point of CaCuSi$_4$O$_{10}$, which was 1064.4\,{\textcelsius}, with the onset of decomposition of this substance occurring at $T=1021.3$\,{\textcelsius}.
The endothermic peak in combination with the mass loss step (0.62\,\%) clearly indicates that the incongruent melting of cuprorivaite is accompanied by partial chemical reduction of copper (Cu$^{2+}$~$\rightarrow$~Cu$^{+}$) with the release of oxygen.
Melting of a mixture of phases and the formation of a complex eutectic involving tridymite and pseudowollastonite takes place at this point.
As a result, cuprorivaite turns into silicate melt rich in copper and calcium as well as CaO\,--\,CuO\,--\,SiO$_2$ eutectic.

The next endothermic peak at $T=1134.9$\,{\textcelsius} is highly likely associated with further melting of the eutectic in the CaO\,--\,CuO\,--\,SiO$_2$ system.
In this case, the decrease in mass by 0.084\,\% can be explained by further loss of oxygen from the melt.

Now, let us consider low-temperature exothermic peaks at 64.7 and 256.2\,{\textcelsius}. 
It should be noted that they are barely visible on the DSC curve in the double-cycle experiment, whereas in the single-cycle one they are clearly visible.%
\footnote{%
This is a result of a smaller sample mass during the latter experiment.
Reducing the sample mass to 4.0~mg allowed the instrument to detect subtle effects such as tridymite peaks that might otherwise be masked when using a more massive sample, as can be seen in the DSC curve of the double-cycle experiment, where these peaks are difficult to distinguish.
With a large sample weight (13~mg), the thermal inertia of the sample is higher. 
Weak low-temperature signals can be ``washed out'' against the baseline or due to the temperature gradient within the massive powder layer.
With a small sample weight (4~mg), the thermal conductivity is higher, and the instrument clearly detects subtle energy transitions in the crystal lattice.%
}
We believe that these peaks correspond to polymorphic transformations of tridymite.
The peak at 256.2\,{\textcelsius} is a classical $\alpha \rightarrow \beta$ tridymite transition (or transition between monoclinic and hexagonal/orthorhombic forms), 
whilst the peak at 64.7\,{\textcelsius} may be interpreted as one of the additional transitions within the complex structure of tridymite (often referred to as a transition between low-temperature modifications $L1 \rightarrow L2$) \cite{Shahid1970, graetsch1991, heaney1994, hirose2005};
the exothermic nature of these transitions in the initial heating cycle suggests the relaxation of structural strains induced during the industrial synthesis of the pigment.


The DSC curve for cooling the light sample after the first heating shows an exo-peak at a temperature of 1096.8\,{\textcelsius}.
This peak represents the liquidus point of the main eutectic CaO\,--\,CuO\,--\,SiO$_2$ of the system.
This means that at 1450\,{\textcelsius} the melt was homogeneous, and crystals, probably tridymite, began to precipitate from the melt at exactly this temperature.

Upon further cooling, the next exo-peak was detected at 1056.0\,{\textcelsius}, likely associated with the onset of precipitation of the most refractory phase, namely pseudowollastonite.
Its melting point in pure form is higher 
\cite{pseudowollastonite_melting, pseudowollastonite_melt, wollast-pseudowollast, pseudowollastanite_meteorites}, but it can drop to this level in copper--silicate melt.

Further cooling resulted in the appearance of the exothermic peaks at 886.3 and 861.5\,{\textcelsius}, which may be assigned to crystallization of tenorite and cuprite, and probably also tridymite.
The splitting into two peaks indicates that two different phases were crystallizing, or one phase was crystallizing in two stages (e.g., first the bulk, then the eutectic).

Thus, after cooling the crucible in the first cycle, a frit was formed with the main crystalline phases of pseudowollastonite, tenorite and tridymite, each in a proportion of about 30 wt.\% (Table~\ref{tab:Phases}), as well as a moderate amount of cuprite ($\sim6$\,wt.\%) and a minor proportion of lime ($\sim1$\,wt.\%). 

Let us move on to examining the second annealing cycle to which the heavier sample was subjected.
The first strong exothermic peak is observed at 835.4\,{\textcelsius} with no change in mass.
We believe that this peak may be due to intense crystallization of the amorphous phase (glass) formed during cooling, i.e. the silicate matrix is being ordered without any change in mass.
The next two endothermic peaks at 977.5 and 1096.8\,{\textcelsius} might be stepwise dissociation of copper oxides within a silicate matrix.
Thus, the peak at 977.5\,{\textcelsius} may be associated with the melting of the local eutectic, which promotes the decomposition of the residues of complex compounds, whilst the temperature of 1096.8\,{\textcelsius} is close to the melting point of the Cu$_2$O\,--\,SiO$_2$ eutectic and/or ultimate reduction of copper (CuO\,$\rightarrow$\,Cu$_2$O). 
Finally, the endo-peak at $T=1134.9$\,{\textcelsius}, as mentioned above for the first heating, might be associated with melting of the eutectic in the CaO\,--\,CuO\,--\,SiO$_2$ system. 

Now, let us estimate the percentage of copper that is reduced during heating and then re-oxidized during cooling.
Since, if all Cu$^{2+}$ contained in cuprorivaite were to be reduced upon heating, the change of mass would be $\Delta m_{\rm s} \approx -2.13$\,\%, but the value of $\Delta m_{\rm s}$ registered in the first cycle was $-2.00$\,\% for the heavier sample and nearly $-2.1(3)$\,\%, for the lighter one, then in excess of 94\,\% of copper (up to 100\,\% for the smaller sample) was converted to the Cu$^{+}$ state at 1450\,{\textcelsius}.

Analogously, after cooling in the first cycle, we obtained $\Delta m_{\rm s} \approx 1.00$\,\% for the heavier sample and nearly $1.5(1)$\,\%, for the lighter one.
Consequently, about 50\,\% of copper for the heavy sample and more than 70\,\% of copper for the light one was re-oxidized to Cu$^{2+}$ due to capture of oxygen from the atmosphere in the crucible that agrees with the percentage of cuprite given in Table~\ref{tab:Phases}.

After the second heating of the heavier sample, $\Delta m_{\rm s} \approx -1.06$\,\%;
this means that the approximately 50\,\% of copper that oxidized during the first cooling reverted to the Cu$^{+}$ state.
Thus, in the second cycle, the copper was again in a virtually completely reduced state (Cu$^{+}$) when heated to 1450\,{\textcelsius}. 
The degree of copper reduction in the melt relative to the original CaCuSi$_4$O$_{10}$ was 98 to 100\,\%.
Cooling down to the room temperature returned almost half of the copper to the Cu$^{2+}$ state again.
Since this occurred over a wide temperature range (1300 to 700\,{\textcelsius}), oxygen slowly diffused into the cooling melt or glass from the atmosphere inside the crucible.

Since only tridymite was detected by XRD after the second cycle of annealing, it can be concluded that all other elements, namely copper (CuO~+~Cu$_2$O) and calcium (CaO), were eventually incorporated into the silicate glass.
Since calcium and copper could not penetrate the tridymite crystal lattice, they formed silicate glass; 
yet to form stable glass, they needed to take some of the silicon dioxide from the system.
Let us estimate the minimum amount of glass formed during the double-cycle annealing. 
The glass incorporated all 15\,\% of CaO and all 21\,\% of CuO contained in cuprorivaite.
During the first cycle of annealing, the product formed contained 32.6\,\% of CaSiO$_3$ (Table~\ref{tab:Phases}), which consists of approximately 48\,\% of CaO and 52\,\% of SiO$_2$.
Therefore, these 15\,\% of CaO retained at least 16\,\% of SiO$_2$ in stoichiometric glass.
Thus, an estimate of the final weight percentages of copper-lime-silicate glass and tridymite based on XRD analysis and mass balance showed that after two annealings of the heavy specimen, a glaze consisting of approximately 52\,wt.\% of glass and 48\,wt.\% of tridymite crystals was formed.

\subsubsection{Polarizing microscopy} \label{subsubsec:PM}

Let us dwell on the PM analysis of the product of the first cycle of annealing.
Acicular crystals found by PM in the samples after the first cycle of annealing (Figs.\ref{fig:PM_02-18_a}\,--\,\ref{fig:PM_02-18_r}), according to their optical properties, exactly correspond to the high-temperature hexagonal tridymite crystals detected by XRD (Table~\ref{tab:Phases}).
Tridymite crystals of acicular habit synthesized because they precipitated from supercooled melt due to relatively high cooling rate of the sample during TA.\footnote{%
Tabular (platy) tridymite crystals are not observed in the preparations \cite{tridymite_mindat}.%
}
The next features, the bright areas exhibiting birefringence and interference patterns in crossed polars, are identified as tabular, mainly platy, pseudowollastonite crystals also revealed by XRD (Table~\ref{tab:Phases}).
Both tridymite and pseudowollastonite crystals are cemented by glass, which in PM images appears as transparent areas in parallel polars and dark areas in crossed ones, as well as dark regions in parallel polars that turn into dark and red or orange domains when viewed under crossed polars.
These regions appear dark, when viewed under parallel polars, owing to strong absorption of light.
The dendritic structures colored black or red are recognized as domains of glass filled, respectively, with crystallites of tenorite or cuprite, which are detected by XRD.
The areas of the photographs taken with crossed polars, which appear red or orange grainy, are also formed by microcrystals of cuprite in glass, whilst the dark regions without a dendritic structure are tenorite crystallites incorporated in glass.
Of course, Cu$_2$O crystallites do not exhibit birefringence, however, they are observed in the images obtained with crossed polarizers.
Apparently, this occurs due to Rayleigh scattering of incident polarized light, as well as light reflected in the glass, on these crystals, which changes its original polarization, as a result of which it passes through the crossed polarizers.
As a matter of fact, the red crystallites of cuprite are visible in the dark field mode, which in this case is implemented due to the crossed polarizer configuration.

Thus, PM analysis has qualitatively proved the indexing of XRD patterns recorded after the first cycle of annealing and clearly demonstrated the presence of pseudowollastonite, tenorite, tridymite and cuprite into the crucible.
This analysis have also shown that tenorite and cuprite crystallites had been incorporated into glass, which had cemented larger high-temperature crystals of pseudowollastonite and tridymite into a conglomerate during melt vitrification. 

Let us focus on the PM analysis of the product of the first annealing cycle.
Like after the first cycle of annealing, it should be concluded that acicular crystals found by PM in the samples after the second cycle of annealing (Figs.~\ref{fig:PM_07-09_a}\,--\,\ref{fig:PM_07-09_b}), according to their optical properties, are recognized as tridymite crystals detected by XRD (Table~\ref{tab:Phases}).
These crystals form frit with green glass.
Since Ca and Cu are components of this glass, its formula is CuO\,--\,Cu$_2$O\,--\,CaO\,--\,SiO$_2$, and because of the non-uniformity of concentration of Cu$^+$, Cu$^{2+}$ and Ca, it slightly changes its color between shades of green. 
Minor amount of nanocrystallites of tenorite was also detected in the crucible, which was not detected by XRD analysis.
Besides, particles of platy crystals, which remained blue when viewed under both parallel and crossed polars, were found.
These particles are small lamellae of cuprorivaite.
This means that some, very small, amount of cuprorivaite still managed to re-synthesize during the cooling of the crucible in the second cycle. 

Thus, the formation of only copper-lime silicate glass and tridymite crystals as a result of two annealings has been confirmed using PM microscopy.

\section{Conclusion} \label{sec:concl}

In conclusion, the main findings of the article should be formulated.

The first and the most important point is that cuprorivaite begins to decompose by incongruent melting at a temperature of about 1020\,{\textcelsius}, with the minimum of the DSC endothermic peak associated with this process being at 1064.4\,{\textcelsius}.
It follows from this that the maximum temperature of synthesis or annealing of any of its physical forms---powder, film, single crystal or polycrystal---should not exceed 1020\,{\textcelsius}.
Moreover, the temperature of any processing of structures, in which it is included, must not exceed this value.

Next, cuprorivaite decomposes irreversibly and subsequent cyclic annealings up to a temperature of 1450\,{\textcelsius} do not lead to its re-synthesis, at least at heating and cooling rates of 20\,{\textcelsius}/min.

Finally, as a result of two successive annealings up to the temperature of 1450\,{\textcelsius} at heating and cooling rates of 20\,{\textcelsius}/min, cuprorivaite is transformed into a two-phase system consisting of acicular crystals of monoclinic tridymite fused with green glass with the composition CuO\,--\,Cu$_2$O\,--\,CaO\,--\,SiO$_2$, with the weight ratio of tridymite to glass being about $12:13$.

 \section*{CRediT authorship contribution statement}

\textbf{Vladimir A. Yuryev:}
Conceptualization,
Methodology,
Funding acquisition, 
Investigation,
Formal analysis,
Writing\,--
Original draft, 
Writing\,-- Review
\& Editing,
Supervision.
\textbf{Sergey~V.~Kuznetsov:}
Methodology,
Investigation,
Writing\,--
Original draft.
\textbf{Alexander~A.~Alexandrov:}
Methodology,
Investigation.
\textbf{Tatyana V. Yuryeva:}
Investigation.


\section*{Declaration of competing interest} 

The authors declare that they have no known competing financial interests or personal relationships that could have appeared to influence the work reported in this paper.

\section*{Data availability} 
The data required to reproduce the findings of this study are available from the corresponding author upon request.


\section*{Acknowledgments}

We thank Mr. Ilya~B.~Afanasyev for the SEM images in Fig.~\ref{fig:SEM+PM}.

The research was accomplished under the non-profit scientific collaboration agreement between GPI RAS and GOSNIIR.
 
We express our gratitude to the Center for Shared Use of Scientific Equipment of GPI RAS 
for supporting this study by providing access to its equipment.

\clearpage


\section*{Figures and Tables}

\clearpage

\begin{table}[htb]
	\caption{%
		DSC peaks and sample mass change $\rm\Delta$$m_{\rm s}$.%
		\label{tab:Peaks}%
	}
	\begin{center}
		{
			\begin{tabular}{cccccc}
				\hline
Experiment type & Cycle No. & Heating$/$cooling & Temperature & Peak type & $\rm\Delta$$m_{\rm s}$ \\ \hline
(double$/$single) &  & ($\uparrow/\downarrow$) & ({\textcelsius}) & (exo$/$endo) & ({\%}) \\ \hline
single & 1 & $\uparrow$ & 64.7 & exo & -- \\
single & 1 & $\uparrow$ & 256.2 & exo & -- \\
double\,\&\,single & 1 & $\uparrow$ & 1064.4 & endo & $-0.62$ \\
double\,\&\,single & $1\,\&\,2$ & $\uparrow$ & 1134.9 & endo & $-0.084$ \\
single & 1 & $\downarrow$ & 1096.8 & exo & -- \\
single & 1 & $\downarrow$ & 1056.0 & exo & -- \\
single & 1 & $\downarrow$ & 886.3 & exo & -- \\
single & 1 & $\downarrow$ & 861.5 & exo & -- \\
double & 1 & $\uparrow$ & -- & -- & $-2.00$ \\
double & 1 & $\downarrow$ & -- & -- & $+1.00$ \\
single & 1 & $\uparrow$ & -- & -- & $-2.1(3)$ \\
single & 1 & $\downarrow$ & -- & -- & $+1.5(1)$ \\
double & 2 & $\uparrow$ & 835.4 & exo & -- \\
double & 2 & $\uparrow$ & 977.5 & endo & $-0.15$ \\
double & 2 & $\uparrow$ & 1096.8 & endo & $-0.11$ \\
double & 2 & $\uparrow$ & -- & -- & $-1.06$ \\
double & 2 & $\downarrow$ & -- & -- & $+0.94$ \\
				\hline
			\end{tabular}
		}
	\end{center}
\end{table}

\clearpage

\begin{table}[htb]
	\caption{%
	Phase composition of the substances, the powder patterns of which are presented in Fig.~\ref{fig:XPA}; the percentage content of the identified crystalline phases was estimated using the reference intensity ratio method.%
	\label{tab:Phases}%
	}
	\begin{center}
		{
			\begin{tabular}{cccccc}
				\hline
				%
				%
				Sample  &Phase name &Formula & Crystal System  &  Database Entry  &  Amount (wt.\%)   \\
				\hline
				Original & Cuprorivaite  & CaCuSi$_4$O$_{10}$ & Tetragonal & PDF 01-085-0158 & 100.0 \\
				After 1 cycle 	& Pseudowollastonite  	& CaSiO$_3$  &  Monoclinic  & PDF 04-012-1764 &  32.6 \\
				 					& Tenorite 	& CuO  &  Monoclinic  & PDF 00-048-1548	&   32.2	\\
				 					& Tridymite 	& SiO$_2$  &  Hexagonal  & PDF 00-018-1169 &  28.3 \\
				 					& Cuprite 	& Cu$_2$O  &  Cubic  & COD 1010941 &  5.9 \\
				 					& Lime 	& CaO  &  Cubic  & COD 9006719 & 1.0 \\
				After 2 cycles & Tridymite	& SiO$_2$ & Monoclinic & PDF 04-008-8461 & 100.0 \\
				\hline
			\end{tabular}
		}
	\end{center}
\end{table}


\end{document}